\documentclass[11pt]{article}
\usepackage{bm,amsmath,color,dsfont, amssymb}
\usepackage[titletoc,toc,title]{appendix}
\usepackage{graphicx}
\usepackage[section]{placeins}
\usepackage{esint}
\usepackage{adjustbox}
\usepackage[hidelinks]{hyperref}
\usepackage{caption}
\usepackage{cleveref}
\usepackage{subcaption}
\usepackage{tabularx}
\usepackage{tgtermes} 
\usepackage{relsize}
\usepackage{booktabs}
\usepackage{wrapfig}
\usepackage{siunitx}
\usepackage{bm}
\usepackage{xcolor}
\usepackage{booktabs}
\usepackage{multirow}
\usepackage{float}

\usepackage{authblk}
\begin{document}
\title{The Self-Organized Criticality Paradigm \\ in Economics \& Finance}
\author{Jean-Philippe Bouchaud \\ Capital Fund Management \& Acad\'emie des Sciences}
\maketitle
\begin{abstract}
``Self-Organised Criticality'' (SOC) is the mechanism by which complex systems are spontaneously driven towards, or even across, a {\it critical point} at the edge between stability and chaos. These special points are characterized by fat-tailed fluctuations and long-memory correlations. Such a scenario can explain why insignificant perturbations may generate large disruptions, through the propagation of ``avalanches'' across the system. In this short review, we discuss how SOC could offer a plausible solution to the excess volatility puzzle in financial markets and the analogue ``small shocks, large business cycle puzzle'' for the economy at large, as initially surmised by Per Bak et al. in 1993 \cite{bak1993aggregate} and, in a different language, by Hyman Minsky. We argue that in general the quest for efficiency and the necessity of {\it resilience} may be mutually incompatible and require specific policy considerations. 
\end{abstract}

\tableofcontents

\section{Introduction}

Many systems made up of a large number of interacting items appear to be ``marginally stable'', i.e. close to an incipient instability, or tipping point. The scenario according to which these systems are spontaneously driven towards such a fragile state is called ``self-organised criticality'' (SOC), and was proposed by Per Bak \cite{bak2013nature} as a generic mechanism that allows one to explain large endogenous fluctuations in complex systems, ranging from natural systems (avalanches, earthquakes, floods, solar flares, turbulence...) \cite{bak1987self, sornette1989self, frisch1995turbulence, sethna2001crackling, sachs2012black, watkins201625}, ecological systems (mass extinctions) \cite{may1972will, bak1997mass}, biological neural networks (epilepsy) \cite{de2006self, chialvo2010emergent, osorio2010epileptic, kinouchi2020mechanisms}, bird flocks and fish schools (collective motion) \cite{bialek2014social}, socio-technical systems (black-outs, traffic jams, failure cascades,...) \cite{dekker2021cascading,laval2023self, Moran2024}, etc., etc. 

Financial markets and global economies are equally prone to such wild fluctuations. History is strewn with bubbles and crashes, booms and busts, crises and upheavals of all sorts. Understanding the origin of these dramatic events (and the remote possibility of curbing them) is arguably one of the most important problems in economic theory. Are these events exogenous (due to a shock from outside the system) or endogenous (generated by internal feedback loops)? \cite{sornette2006endogenous}. As Cochrane quipped \cite{cochrane1994shocks}, {\it What shocks are responsible for economic fluctuations? Despite at least two hundred years in which economists have observed fluctuations in economic activity, we still are not sure} (!). 

This did not escape Per Bak's sagacity. Together with Kan Chen, José Scheinkman and Michael Woodford \cite{bak1993aggregate, scheinkman1994self}, he was quick to transpose ideas born in the context of statistical physics to economic situations, with the suggestion that the substantial, unexplained year on year GDP fluctuations of large developed economies -- the so-called ``small shocks, large business cycle puzzle'' \cite{Bernanke1996} -- could in fact result from the {\it inherent fragility} of economic systems that prevents fluctuations to vanish, even for large system sizes.  

The financial markets analogue of excess GDP fluctuations is the equally well-known ``excess volatility puzzle'' elicited by R. Shiller \cite{shiller1981stock, shiller1987volatility} and by LeRoy-Porter \cite{leroy1981present}.\footnote{For a nice summary of the literature on this point, see \cite{leroy2006excess}.} Asset prices frequently undergo large jumps for no particular reason, when financial economics asserts that only unexpected news can move prices \cite{cutler1988moves, joulin2008, marcaccioli2022exogenous}. Volatility is an intermittent, scale invariant process that resembles the velocity field in turbulent flows \cite{frisch1995turbulence,muzy2000modelling}. Small, seemingly innocuous perturbations can end up sending the market in shambles -- like ``Black Monday'' in October 1987 or the infamous ``Flash Crash'' of May 6th, 2010. But such extreme events are not isolated outliers. Just as earthquake severity can span orders of magnitude, from hardly detectable tremors to devastating calamities, the probability distribution of price returns exhibit a power-law tail typical of complex systems sitting in the vicinity of a critical point (see e.g. \cite{cont2001empirical,bouchaud2003theory, gabaix2009power}). Again, temptation is high to invoke a kind of spontaneous coordination of competing market participants right at the border of chaos. 

The aim of this (short) review is to discuss whether such a paradigm makes sense in the context of economic systems and financial markets. Indeed, as the French saying goes, {\it comparaison n'est pas raison}. It may well be that all these analogies are misleading, and that the true underlying mechanisms leading to excess volatility should be found elsewhere. For one thing, some authors would still argue that volatility is not excessive at all, but merely reflects the existence of genuine exogenous shocks of large amplitude, perhaps invisible to the economist, but adequately processed by smart market participants and economic agents. Needless to say, such an ``invisible hand'' explanation often sounds preposterous, but reflects that it is hard to accept (and quite disturbing indeed) that large events can occur without ``large causes'' -- whereas it is the major epistemological lesson of complexity science in general, and SOC in particular, that tiny perturbations can induce full crises. A complex system can actually be defined as a system where small perturbations {\it can} -- but not necessarily {\it do} -- trigger incommensurate effects, see e.g. \cite{parisi2007physics, PARISI1999557, sethna2001crackling, bouchaud2021radical}.

Other explanations have been proposed, which are more plausible than phantom shocks. For example, the ``Granularity Hypothesis'' (GH) of X. Gabaix \cite{gabaix2011granular} is an attempt to rationalize the fact that idiosyncratic shocks may survive at the aggregate level, because some firms have a disproportionate size and cannot be averaged out (see also \cite{moran2024revisiting}). This is the case when firm sizes $S$ are power-law distributed, with a p.d.f. tail decaying slow enough, for example as $S^{-2}$ as is indeed the case empirically \cite{axtell2001zipf}. Similarly, according to \cite{gabaix2003theory}, large price jumps in financial markets would result from orders sent by large investors or funds, the size of which is also known to be heavy-tailed \cite{gabaix2006institutional} (but probably not power-law tailed, see \cite{schwarzkopf2010empirical}). 

Although superficially very different, SOC and GH share common ingredients. Whereas the Granularity Hypothesis ascribes the persistence of aggregate fluctuations to large, stable entities (firms, investors), the SOC scenario relies on large, fleeting ``coalitions'' that are dynamically generated by contagion across the system. In other words, the survival of idiosyncratic shocks at the aggregate level is a result of their propagation over large scales -- a phenomenon made possible by {\it fragility}, i.e. the proximity of a critical point. While large coalitions or clusters are temporary in the SOC scenario, they are persistent within GH. 

Yet another mechanism, that we will discuss further in sections \ref{sec:stabilizing} and \ref{sec:firm_ecology_2}, is that the equilibrium state of the system is in fact dynamically unstable, leading to a quasi-periodic or chaotic evolution even {in the absence} of any exogenous shock. Excess volatility is then purely self-induced. As illustrated by the model of section \ref{sec:firm_ecology_2}, one does not necessarily expect to see power-law statistics or long-memory effects when such systems are left on their own device. Interestingly, however, attempts to stabilize the system (through e.g. regulation, monetary policy, learning, etc.) might yet again drive the system close to a critical point, reinstalling, in a sense, the SOC scenario. 

Finally, let us mention two related but different concepts that we will not develop further here: Highly Optimized Tolerance (HOT) \cite{carlson1999highly} and Self-Organized Bistability (SOB) \cite{di2016self}, which have been argued to be relevant in different situations. Such scenarios  might be also interesting to consider in the context of economic or financial systems, see e.g. \cite{harras2012noise} for similar ideas, and section \ref{sec:conclusion}.

The outline of this paper is as follows. We first explain in section \ref{sec:simple} what is special about critical (marginally unstable) points. We then motivate the concept of Self-Organized Criticality in section \ref{sec:SOC} and give a few examples where such a concept naturally applies. We next turn to possible applications of SOC in economical (section \ref{sec:SOC_E}) and financial (section \ref{sec:SOC_F}) contexts. Finally, we discuss in the conclusion section \ref{sec:conclusion} the policy implications of the possible fragility of socio-economic systems. We argue in particular that efficiency and resilience are often incompatible, and that operators, regulators and policy makers should take stock of the unintended consequences that can appear when {\it resilience} (i.e. tolerance to tail events) is not explicitly included in the welfare function. 

\newpage

\section{Stability and Criticality}\label{sec:simple}

\subsection{A trivial example}\label{sec:trivial}

Consider a simple dynamical equation for a quantity $x(t)$ that describes deviations from equilibrium, for example the difference between the actual production of a firm and the long term equilibrium (optimal) value. We write:
\begin{equation} \label{eq:OU}
    \frac{{\rm d} x}{{\rm d}t} = - \kappa x(t) + \eta(t),
\end{equation} 
where $\kappa$ is the anchoring parameter, which has dimensions of inverse-time, and $\eta(t)$ is a Gaussian white noise describing unanticipated exogenous shocks, with $\mathbb{E}[\eta(t)]=0$ and $\mathbb{E}[\eta(t) \eta(t')]= \sigma^2 \delta(t-t')$. This equation defines an Ornstein-Uhlenbeck process.

In the absence of noise, $\sigma=0$ and $x(t)=x_0 \exp(-\kappa t)$, i.e. the system relaxes to equilibrium in a time $t_{\rm{eq.}} \sim \kappa^{-1}$. In the presence of noise, $x(t)$ randomly fluctuates around zero, with a stationary correlation function given by
\begin{equation}
   \lim_{\kappa t \gg 1} \mathbb{E}[x(t+\tau) x(t)] = \frac{\sigma^2}{2\kappa} e^{-\kappa \tau},
\end{equation}
with in particular $\mathbb{E}[x^2] = { \sigma^2}/{2\kappa} \propto t_{\rm{eq.}}$. The important message of this ultra-simple model is that as equilibrium becomes unstable, i.e. $\kappa \to 0$, the variance of fluctuations and the relaxation time both diverge at the same rate, as $\kappa^{-1}$. 

A slightly less trivial example is the multidimensional generalisation of Eq. \eqref{eq:OU}, namely:
\begin{equation} \label{eq:OU_d}
    \frac{{\rm d}x_i}{{\rm d}t} = - \sum_{ij} \mathbb{K}_{ij} x_j(t) + \eta_i(t), \qquad i,j = 1, \ldots, N
\end{equation} 
where $\mathbb{K}$ is the so-called stability matrix and $\eta_i(t)$ is a multidimensional white noise. Now, equilibrium stability depends on the eigenvalues of matrix $\mathbb{K}$, which are in general complex. If all eigenvalues have negative real part, equilibrium is stable. Let us denote by $-\kappa^\star$ the real part of the eigenvalue closest to zero. Then one finds that generically, when $\kappa^\star \to 0$, the variance of the fluctuations of any component $\mathbb{E}[x_i^2]$ diverges as $(\kappa^{\star})^{-1}$, as does the relaxation time of the system. 

Hence, the main message is that in the limit of marginal stability $\kappa^\star \to 0$, the system both amplifies exogenous shocks and becomes auto-correlated over very long time scales.   

\subsection{Critical Branching}\label{sec:branching}

A less trivial, yet classic example, is the critical branching transition. The model describes a very large class of situations: sand pile avalanches, brain activity, epidemic propagation, default/bankruptcy waves, word of mouth,  etc. In the language of sand piles, one assumes that one rolling grain can dislodge a certain number $n$ of other grains that start rolling and may themselves dislodge more grains downhill. Suppose that $n$ is an IID random variable with distribution $\rho(n;R_0)$ with a finite second moment and $\mathbb{E}[n]=R_0$. 

When $R_0 < 1$, a first unstable grain dislodges on average a finite number of other grains. More precisely, the average size of avalanches is given by $(1-R_0)^{-1}$. The probability of very large avalanches is exponentially small. However, as $R_0$ approaches $1$, one single grain can lead to large disruptions. The probability of triggering an avalanche involving $S \gg 1$ grains is given by \cite{harris1963theory}
\begin{equation} \label{eq:avalanches}
    P(S) \propto S^{-3/2} \exp(-\varepsilon^2 S), \quad \qquad \varepsilon = 1 - R_0 \to 0.
\end{equation}
In other words, when $R_0=1$ the distribution of avalanche sizes is a scale-free, power-law distribution $S^{-3/2}$, with infinite mean.\footnote{Note that when the distribution of off-springs $\rho(n;R_0)$ decays as $n^{-1-\alpha}$ with $1 < \alpha < 2$, the critical avalanche size distribution has a power-law tail given by $S^{-1-1/\alpha}$. Note that for $\alpha=2$ one recovers the classic $S^{-3/2}$ behaviour. } Still, the probability $\phi$ that the avalanche continues forever is zero. 

When $R_0 > 1$, this probability becomes non-zero (and grows as $\phi \propto R_0 - 1$ when $R_0$ is close to unity). With probability $\phi$, the size of the avalanche is formally infinite, and in practice only limited by the size of the sand pile. With probability $1-\phi$, the avalanche stops and its size is  again distributed as in Eq. \eqref{eq:avalanches}. 

The value $R_0=1$ is therefore a {\it critical point}, separating a stable regime with avalanches of finite size from an unstable regime where landslides occur endogenously, triggered by the initial motion of a single grain. As for the trivial example of the previous section, such an instability is accompanied by a diverging time scale. Indeed, the duration $\tau$ of an avalanche is related its size as $S \propto \tau^2$ \cite{harris1963theory}. Correspondingly, durations are distributed again as a power-law ($\propto \tau^{-2}$) for $R_0=1$.

Note that we have called $R_0$ the mean number of off-springs $\mathbb{E}[n]$ by analogy with the notation used for the basic reproduction number in epidemics, with a rather obvious mapping. When infected individuals transfer the virus to $R_0 > 1$ (on average) other individuals, there is a non zero probability $\phi$ for a full-blown pandemic to spread across the population.  

\section{Self-Organized Criticality}\label{sec:SOC}

\subsection{Motivation}\label{sec:motivation}

The previous example shows that close to criticality, a system can reveal a very interesting type of phenomenology: when $R_0 \lesssim 1$, small shocks {\it can} generate large perturbations, but do not necessarily do so. In fact, the critical power-law distribution $S^{-3/2}$ means that most ``avalanches'' are of small size, although {\it some} can be very large. In other words, the system looks stable, but occasionally goes haywire with no apparent cause. 

However, this only occurs for a very special value of $R_0$. In generic cases, the system is either stable, or completely unstable. Why would the proximity of $R_0=1$ play a role in practice, except if the system is fine-tuned ``by hand'' close to criticality? Why would criticality be relevant in practice, in particular in the context of economics and finance?

The seminal idea of Per Bak is to think of {\it model parameters themselves as dynamical variables}, in such a way that the system spontaneously evolves towards the critical point, or at least visits its neighbourhood frequently enough (for a precursor of this general idea, see section 8 of Ref. \cite{keeler1986robust}, and for reviews, see \cite{bak2013nature, jensen1998self, watkins201625}). Take for example the case of highly transmissible virus, with a natural value of $R_0$ above unity. As the disease spreads, containment measures are adopted (masks, social distancing, etc.) and the risk of getting sick is taken seriously by the population. This leads to a significant reduction of $R_0$, perhaps even below unity. Now, the epidemics slows down and people become more complacent, which causes $R_0$ to rise again. It is not absurd to imagine that the process will indeed self-organize around a point where the spread of the disease is curbed at a minimal amount of individual constraints. This is achieved precisely at the critical value $R_0=1$. 

\subsection{Sweeping through an instability }\label{sec:sweeping}

Consider now the sand pile example initially put forth by Per Bak and collaborators \cite{bak1987self}. Start with a flat layer of sand and slowly add grains from a point source above it. A conical pile will form, with a progressively steeper slope. As the slope increases, so does the probability $R_0$ that a rolling grain destabilizes more grains increases. We will thus describe the dynamics of the pile in terms of $R_0$ rather than its slope.

As long as $R_0 < 1$, the resulting avalanches are small and do not change the overall shape of the pile. But as soon as $R_0 > 1$, there is a possibility of a landslide that collapses the pile and reduces its slope. The dynamical evolution of the probability to find a pile with ``slope'' $R_0$ is then described as\footnote{Throughout the paper we use the standard notation $(x)^+ = \max(x,0)$.}
\begin{equation}\label{eq:slope}
    \frac{\partial P(R_0,t)}{\partial t} = - \mu  \frac{\partial P(R_0,t)}{\partial R_0} - \gamma (R_0 - 1)^+ P(R_0,t),
\end{equation}
where $P(R_0,t)$ is the probability to find a slope characterized by a certain value of $R_0$, $\mu$ describes the rate at which $R_0$ increases due to the addition of new grains, and $\gamma$ the rate at which single grains start rolling down the slope, and $(R_0-1)^+$ the probability that a single grain does trigger a landslide. For simplicity, one assumes that after the landslide, the pile restarts at $R_0=0$. 

\begin{figure}[h!]
    \centering
    \includegraphics[width = 0.5\textwidth]{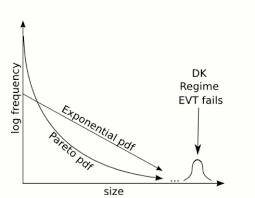}
    \caption{\small{Schematic illustration of the concept of Dragon Kings (DK). Plot of the distribution of avalanche sizes $S$ (say) which follow a Pareto law for sizes $S$ much less than the maximum possible size $S_{\max}$, followed by an anomalous hump around $S_{\max}$ (corresponding to system-wide avalanches). The mechanism generating those Dragon Kings is, in many cases, the excursion of the system in an unstable region, whereas the power-law regime comes from the vicinity of the critical point. From https://encyclopedia.pub/entry/30734
} }
    \label{fig:DK}
\end{figure}

The stationary state $Q_{\text{st.}}(R_0) = P(R_0,t \to \infty)$ of this process reads: 
\begin{align} \nonumber
    Q_{\text{st.}}(R_0) &= Z^{-1}  &(0 \leq R_0 \leq 1),
    \\ &= Z^{-1}  \exp\left(-\frac{\gamma}{2 \mu} (R_0-1)^2\right)  &(R_0 \geq 1),
\end{align}
with $Z=1+\sqrt{{\pi \mu}/{2\gamma}}$ such that $Q_{\text{st.}}(R_0)$ is normalized to one. This equation shows that there is a finite probability density for the slope to be exactly critical. This is enough to ensure that one will observe a power-law distribution of avalanche sizes \cite{sornette1994sweeping}, which can be computed from 
\[
\mathcal{P}(S) = \int_0^\infty {\rm d}R_0 \, Q_{\text{st.}}(R_0) P(S),
\]
where $P(S)$ is given by Eq. \eqref{eq:avalanches}. It is not very difficult to check that $\mathcal{P}(S)$ behaves at large $S$ as $S^{-2}$ without the exponential truncation term present in Eq. \eqref{eq:avalanches}, i.e. as a pure power-law with exponent $-2$ instead of $-3/2$. Note however that $\mathcal{P}(S)$ only captures avalanches of finite size (i.e. much smaller than the total size of the system). One should also consider system-wide avalanches which relax the system all the way down to $R_0=0$. These contribute to an additional hump for very large $S$ of the order of the system size -- events that were called ``Dragon Kings" in \cite{sornette2012dragon} (see Fig. \ref{fig:DK}).

The above example illustrates one possible scenario through which intermittent dynamics, with a power-law distribution of event sizes, can occur (see also \cite{keeler1986robust} for an early discussion). The critical point is regularly visited by the natural dynamics of the system, as it travels from the stable phase to the unstable phase, the instability driving the system back to the stable phase. As discussed below, such a scenario is also reminiscent of the Minsky cycle in financial markets. 

\subsection{Dynamical convergence towards criticality}
\label{sec:GLV} 

A different scenario is when the natural dynamics stops precisely at the critical point. For example, gradient descent applied to complex {\it Constrained Satisfaction Problems} generically stops when a local minimum of the loss function is reached. In many cases however, such minima are marginally stable, that is, the Hessian of the loss function has all its eigenvalues non negative, but the smallest one is very close to zero, see e.g. \cite{cavagna1998stationary,muller2015marginal, Franz2017}. This is precisely the situation described by Eq. \eqref{eq:OU_d}, where stability matrix $\mathbb{K}$ is the Hessian matrix. Therefore, small shocks get amplified by the proximity of an instability and small static perturbations can disproportionately modify the equilibrium state (see also \cite{carlson1999highly} for related ideas). In fact, such perturbations can even change the sign of the smallest eigenvalue, forcing the system to rearrange and find another (marginally stable) equilibrium, often very different from the original (unperturbed) one.

As an illustration, let us consider the so-called Generalized Lotka-Volterra model, describing the population dynamics of many interacting species \cite{bunin2017ecological}. We denote $x_i(t)$ the number of living individuals of species $i=1, \ldots,N$ and assume that these numbers evolve as
   \begin{equation} \label{eq:GLV}
    \frac{{\rm d}x_i}{{\rm d}t} = x_i \left(\mu_i + \sum_{j=1}^N \mathbb{A}_{ij} x_j\right),
\end{equation}  
where $\mu_i > 0$ is the ``fitness'' of species $i$ (i.e. its growth rate in the absence of interaction with other species, including itself) and the matrix $\mathbb{A}$ captures beneficial ($\mathbb{A}_{ij} > 0$) or detrimental ($\mathbb{A}_{ij} < 0$) interactions with other species. (For example, $\mathbb{A}_{ij} \mathbb{A}_{ji} < 0$ describes a predator-prey situation). 

The ecological equilibria corresponding to Eq. \eqref{eq:GLV} must be such that either $x_i^\star=0$ (extinct specie) or $\mu_i + \sum_{j=1}^N \mathbb{A}_{ij} x_j^\star = 0$. 
Naively solving the second equation yields, in a vector notation
\begin{equation} \label{eq:GLV_eq}
  \vec x^\star = - \mathbb{A}^{-1} \vec \mu.  
\end{equation}
However, the resulting vector $\vec x^\star$ will in general contain negative entries, which of course does not make sense since the number of living individuals must be positive or zero. Feasible equilibria are thus such that only a subset of species survives, such that Eq. \eqref{eq:GLV_eq} restricted to that subset returns a vector $\vec x^\star$ with only positive entries. As first anticipated by the late Lord Robert May, not all ecologies can be stable \cite{may1972will}.

\begin{figure}[h!]
    \centering
    \includegraphics[width = 0.5\textwidth]{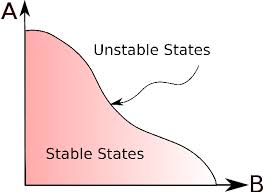}
    \caption{\small{Schematic explanation of the prevalence of marginally stable states in complex systems. One starts from an unstable point (white region) and follows the arrow. The system stops as soon as it becomes stable, i.e. at the boundary between stability and instability. From Ref. \cite{muller2015marginal}.} }
    \label{fig:marginal}
\end{figure}

What about the dynamics generated by Eq. \eqref{eq:GLV}? In the special case where the matrix $\mathbb{A}$ is symmetric, one can show that generically the system evolves towards a marginally stable state, i.e. enough species disappear for a feasible equilibrium to exist, but the sensitivity matrix defined as 
\[
\frac{\partial x_i^\star}{\partial \mu_j} = - (\mathbb{A}^{-1})_{ij}, 
\]
has a diverging eigenvalue $\lambda^\star$ in the large $N$ limit \cite{biroli2018marginally}. This means that any small change in the fitness of one species can have dramatic consequences on the whole system -- in the present case, mass extinctions \cite{bak1997mass}. Similarly, the stability matrix $\mathbb{K}$, given by (cf. Eq. \eqref{eq:OU_d})
\[ 
\mathbb{K}_{ij} = - x_i^\star \mathbb{A}_{ij}
\] 
has an eigenvalue $\kappa^\star \propto \lambda^{\star -1}$ very close to zero, meaning again that one is in a situation where small exogenous shocks are amplified by marginal stability -- see e.g. \cite{stone2018feasibility}. The Generalized Lotka-Volterra equation therefore provides a non-trivial realisation of Eq. \eqref{eq:OU_d} with $\kappa^\star \to 0$, as a natural consequence of the dynamics of the system that converges to a marginally stable equilibrium (see Fig. \ref{fig:marginal}).  We will discuss in section \ref{sec:firm_ecology_1} a direct analogy between Lotka-Volterra dynamics and production networks. 

There are many other concrete examples where such a scenario is at play, see e.g. \cite{muller2015marginal, aspelmeier2019realizable, patil2025emergent}, in particular in the context of complex learning games \cite{PhysRevX.14.021039}. As mentioned above, most complex optimisation systems are, in a sense, fragile, as the solution to the optimisation problem is highly sensitive to the precise value of the parameters of the specific instance one wants to solve, like the $\mathbb{A}_{ij}$ entries in the Lotka-Volterra model. Small changes of these parameters can completely upend the structure of the optimal state, and trigger large scale rearrangements, see e.g. \cite{fisher1991directed, aspelmeier2008bond, krzkakala2005disorder} and, in the context of optimal portfolio construction, Ref. \cite{garnier2021new}.

\subsection{Stabilizing an unstable equilibrium}\label{sec:stabilizing}

Another interesting generic scenario for self-organized criticality was proposed by F. Patzelt and K. Pawelzik in 2011 \cite{patzelt2011criticality}. The narrative is that of the balancing stick problem \cite{cabrera2004human, cabrera2012stick}. Imagine trying to keep vertical a long enough stick on the tip of your finger. The angle with the vertical is noted $\theta(t)$. The vertical position $\theta=0$ is an unstable equilibrium, so the dynamics for small $\theta$ reads
\[
\frac{{\rm d}\theta}{{\rm d}t} = G(t) \theta(t) + \eta(t),
\]
where $G(t) > 0$ is the amplification rate, possibly time dependent, and $\eta(t)$ is a white noise describing unexpected perturbations (e.g. wind). Because $G(t) > 0$, the dynamics is unstable and without any control the stick just falls off. 

The observer will try to maintain the stick vertical by moving their hand in response to the motion of the stick. However, the observation of $\theta(t)$ is usually both noisy (measurement noise) and time-delayed. The observer can therefore only compute their {\it control policy} $F\left(t, \{\theta(t' \leq t - \tau)\}\right)$ with a certain lag $\tau$. The equation of motion then becomes   
\[
\frac{{\rm d}\theta}{{\rm d}t} = G(t) \theta(t) + \eta(t) - F\left(t, \{\theta(t' \leq t - \tau)\}\right).
\]
The optimal control policy is computed in \cite{patzelt2011criticality}. The observer can indeed stabilize the position of the stick around $\theta=0$, but the time series of $\theta(t)$ shows very interesting features (intermittent dynamics, power-law distributions), which are indeed observed in real balancing stick experiments \cite{cabrera2004human, cabrera2012stick} and reminiscent of returns in financial markets \cite{cont2001empirical,bouchaud2003theory, patzelt2013inherent}.

What is the underlying mechanism leading to such special dynamical features? The basic idea here is that the closer $\theta(t)$ is to zero, the more uncertain is the most likely estimate of the amplification rate $\hat G(t) = \dot \theta/\theta$ and the larger the error in the control policy $F(.)$. 

In other words, the better one is able to stabilize the system, the more difficult it becomes to predict its future evolution! This in turn generates occasional large errors, leading to large swings that are later corrected as the control policy becomes efficient again \cite{patzelt2011criticality}. This rather beautiful scenario is quite universal -- for example heart-beat regulation is a result of two conflicting forces: excitation vs. inhibition (sympathetic vs. parasympathetic nervous system), or brain activity regulation \cite{lombardi2017balance}. Such ideas are probably also relevant in economics and finance (see section \ref{sec:liquidity}, \ref{sec:financial_contagion} below). In fact, the title of the present subsection is tantalizingly close to that of Minsky's seminal book, ``Stabilizing an Unstable Economy'' \cite{minsky2008stabilizing}.  

\section{Self-Organized Criticality in Economics}\label{sec:SOC_E}

As mentioned in the introduction, the first attempt to apply SOC to economic systems is due to Bak, Chen, Scheinkman and Woodford in 1993 \cite{bak1993aggregate}. Although their paper is brimming with exciting ideas, the actual model the authors put forth to describe spontaneous large scale output fluctuations appears somewhat ad-hoc. The firm input-output network is modeled as a square lattice, downstream firms buying exactly zero or one unit of good from exactly two upstream firms. At each time step, firms can only produce 2 units of goods or none at all, depending on their inventory and the orders received from their downstream clients. These orders in turn depend on how much these clients need to produce. Demand fluctuations for final goods (i.e. the top row of the lattice) are exogenous, but propagate upstream as ``avalanches'' with a power-law distribution of sizes, like grains on the slope of a sand pile. Hence, again, small local shocks can lead to large scale disruptions of the supply chain. The problem with such a highly stylized model is that it is not clear how generic the results are. In fact, the somewhat constrained production rules have been chosen such that the system is critical -- but, somewhat paradoxically in view of the title of the paper, it is not clear how supply chains in general would dynamically ``self-organize'' into criticality. 

The aim of the present section is to revisit the issue and discuss more recent proposals that lead to self-organized criticality in economic systems, based on arguably less artificial assumptions. The hope is that putting the SOC scenario on firmer theoretical ground could revive the interest of economists. It is indeed fair to say that despite the high profile of its authors, the 1993 paper of Bak et al. \cite{bak1993aggregate} did not reach the level of attention that (in my view) it deserved \footnote{At the time of writing, the joint number of citations of Bak et al. 1993 \cite{bak1993aggregate} and Scheinkman-Woodford 1994 \cite{scheinkman1994self} is $\sim 800$ according to Google Scholar, compared for example to 2,050 for Gabaix' 2011``granularity'' paper
\cite{gabaix2011granular}, or 2,700 for Acemoglu, Carvalho et al. 2012 paper on the network origin of aggregate fluctuations \cite{acemoglu2012network}.}
-- probably because of its purely conceptual nature, and the lack of precise, actionable proposal to fit on data.

\subsection{Macroeconomic fluctuations and ``timeliness criticality''}\label{sec:timeliness}

As a possibly more realistic incarnation of the scenario proposed by Bak et al., let us consider the propagation of production delays along the supply chain. The idea is that when input goods are not timely, production stops -- unless ``buffers'' (i.e. inventories) are present. The same logic holds in other socio-technical systems, such as train or plane networks \cite{dekker2021cascading}. Indeed, the next train can only leave on time if the previous train has arrived and the driver and train managers are available. In this case, time buffers are embedded within the time table, allowing for enough time between different scheduled events, such that delays can be (to some extent) absorbed.   

Recently, Moran et al. \cite{Moran2024} proposed a minimal, stylized model of delay propagation that reveals the existence of a critical point as the size of mitigating buffers is reduced. In a nutshell, the model assumes that the delay $\tau_{i}(n)$ accumulated on node $i$ of the network at the $n^{\rm th}$ iteration step evolves as 
\begin{equation} \label{eq:delays}
    \tau_{i}(n+1) = \left[\max_{j \in \partial_i} \left(\tau_j(n)\right) - B \right]^+ + \epsilon_i(n+1),
\end{equation}
where $\partial_i$ denotes the set of tasks that need to be completed for task $i$ to start, $B$ is the time buffer and $\epsilon_i(n+1)$ is a positive noise describing idiosyncratic events that delay the start of task $i$. The interpretation of the first term in the right hand side of Eq. \eqref{eq:delays} is quite transparent: provided the worst delay is less than the buffer, there is no impact of previously accumulated delay on the start of the new task.  

Above a certain critical buffer size $B_c$, delays are found to self-heal and large-scale system-wide delays are avoided. On the contrary, when temporal margins are not wide enough, delays accumulate without bounds, creating system-wide disruptions. Close to the critical point, delay ``avalanches" of all sizes are observed \cite{Moran2024}. 

Now, there exist a variety of incentives for operators, often reinforced by competitive pressures, to increase time-efficiencies (i.e. reduce $B$) in order to achieve superior operational results. For example, train operators may have the goal to maximize the number of passengers to be transported by the network. This reduction will have a minor impact of the global functioning of the system provided $B$ is larger than $B_c$ -- only small, occasional disruptions ensue. Operators then feel legitimate to reduce $B$ further, until a major breakdown occurs. Taking stock of the recent events, regulators usually step in and impose that reasonable safeguards are put in place. Such a scenario is very close to a sand pile sweeping through its critical angle (see section \ref{sec:sweeping}) or the dynamics of disease propagation, with a reproduction number hovering around $R_0=1$ (section \ref{sec:motivation}). 

The model can be naturally extended to the propagation of supply fluctuations in production networks, where firms reduce costs by keeping the inventory of production inputs at a minimal level. Any delay in such a ``just-in-time'' supply chain can then propagate down the production network with major disruptions, as exemplified by the Suez canal obstruction in 2021. Inventories are conceptualized as buffers that allow firms to keep producing in the absence of inputs for some amount of time. Low levels of inventories can thus cause strong output fluctuations at the aggregate level, see e.g. \cite{Colon2017}. Myopic cost-cutting measures at the local scale may  unwittingly push the whole system into criticality. Such a mechanism provides a clear justification for the seminal idea of Bak et al. \cite{bak1993aggregate, scheinkman1994self} and is perhaps key to explain why large economies are so much more volatile than expected based on economic equilibrium models. 

It is of course tempting to think of other situations in similar terms. For example, bank regulators impose a certain level of capital ``cushions'' (the analogue of buffers) that bank should keep aside to avoid the propagation of liquidity crises. This obviously comes at a cost, since a certain amount of capital cannot be put to work in profitable (but risky) investments. In relatively stable economic environments, bankers will sooner or later argue that safeguards need to be reduced...until the all but inevitable ``Minsky moment'' \cite{minsky} where the default of a single financial institution avalanches through the whole sector. Network models have indeed been proposed to explain the system-wide breakdowns of the banking sector in 2008~\cite{haldane2011systemic,gai2010contagion,squartini2013early, caccioli2018network}.

\subsection{The ecology of production  networks (statics)} \label{sec:firm_ecology_1}

The importance of network effects for the propagation of shocks across production chains was also considered in the context of {\it economic  equilibrium} models in \cite{acemoglu2012network, carvalho}, elaborating on the classic paper of Long and Plosser \cite{long1983real}. Assuming a Cobb-Douglas production function for all firms (see Eq. \eqref{eq:def_ces} below), one can show that idiosyncratic productivity shocks get amplified at the aggregate level, through the existence of client-supplier links between firms. In fact, if the input-output network is such that the degree distribution is broad enough, i.e. if some firms act as ``hubs'' in the supply network, then the volatility of aggregate production decreases slowly (or even not at all) with the number of firms \cite{acemoglu2012network}, possibly explaining why the activity of large economies fluctuate so much.

However, if one drills down further into the mathematics of the model, one realizes that these anomalous fluctuations are in fact due to Gabaix' ``granularity'' mechanism \cite{gabaix2011granular}, namely, when the topology of the network induces a broad distribution of firm sizes (measured as total sales). The idiosyncratic shocks hitting those large firms then overwhelmingly contribute to fluctuations of GDP. A statistical model explaining how the production network might self-organize to generate such a power-law distribution of firm sizes is discussed in \cite{atalay2011network}. 

There are however important issues with equilibrium models. First and foremost, it sounds more plausible to think that the large aggregate disruptions we are trying to decipher are {\it disequilibrium} effects. Indeed, the assumption that during such periods all markets instantaneously clear (no stock-outs, no inventories) and firms muddle through without defaulting is clearly untenable. Second, even if equilibrium is assumed, the choice of a Cobb-Douglas production function is extremely special, as it allows a feasible equilibrium (where all firms produce a positive amount of goods and sell them at positive prices) to exist for arbitrary input-output networks and any value of firm productivities. This is a result of the relatively high amount of substitutability of input goods implied by Cobb-Douglas. As soon as elasticity of substitution $\sigma$ is lower, some firms may have to disappear for a feasible equilibrium to exist -- much as what happens in the context of the Lotka-Volterra description of ecological communities discussed in section \ref{sec:GLV}. Hence, bankruptcies and bankruptcy waves must be considered within such generalized models, which means that a fully out-of-equilibrium, dynamical description of shock propagation is necessary to describe the resulting aggregate fluctuations. 

For definiteness, let us consider the Constant Elasticity of Substitution (CES) family of production functions. Calling $\pi_i$ the production of firm $i$, one writes
\begin{equation}\label{eq:def_ces}
\pi_i = z_i \left(\sum_j a_{ij} \left(\frac{J_{ij}}{Q_{ij}}\right)^{\frac{1}{q}}\right)^{-q} \quad \mbox{with} \quad \sum_{j} a_{ij}=1,
\end{equation}
where $z_i$ is the firm productivity, $Q_{ij}$ is the number of goods firm $i$ buys from firm $j$ and $J_{ij}, a_{ij} \geq 0$ are weight parameters. 

Parameter $q$ captures the substitutability of input goods, and is related to the standard elasticity of substitution through $\sigma=q/(q+1)$. The Cobb-Douglas function $\pi_i= z_i \prod_j (Q_{ij}/J_{ij})^{a_{ij}}$ corresponds to $q \to \infty,\, \sigma=1$ and is often used to describe the average aggregate production of economic sectors~\cite{long1983real}, or of the economy as a whole. When $q \to 0^+,\, \sigma \to 0$, on the other hand, no substitutes are available and Eq. (\ref{eq:def_ces}) reduces to the classical Leontief production function:  $\pi_i = z_i \min_{j} \left({Q_{ij}}/{J_{ij}}\right)$.
It models a situation where redundancy is costly. Firms therefore choose their suppliers with parsimony and cannot ``rewire'' (i.e. find alternative suppliers) on short time scales in the real economy. 

In the case of a general CES production function, the competitive, market clearing equilibrium equation for prices reads \cite{moran2019may}:
\begin{equation}\label{eq:excess_prof2}
\left(z_i p_i\right)^{\zeta} - \sum_{j\neq 0} a_{ij}^{q\zeta} \left(J_{ij} p_j\right)^{\zeta} = a_{i0}^{q\zeta} \left(J_{i0} w_i\right)^{\zeta} \quad (> 0), \qquad \zeta:=\frac{1}{1+q},
\end{equation}
where good $0$ is labor and $w_i$ is the wage of firm $i$. A similar equation can be written for equilibrium productions $\pi_i$. 

Now, in order for the equilibrium to make sense, the solutions to Eq.  \eqref{eq:excess_prof2} must be such that $p_i > 0$, $\forall i$; i.e. that equilibrium prices and quantities must all be strictly positive -- otherwise some firms are for all purposes bankrupt. As first noted by Hawkins \& Simon~\cite{hawkins1949note}, this is {\it not} automatic and requires matrix $\mathbb{M}$, defined by $\mathbb{M}_{ij}= z_i^\zeta \delta_{ij}- a_{ij}^{q \zeta} J_{ij}^\zeta$ to be a so-called ``M-matrix''\footnote{Note that if $\mathbb{M}$ is an M-matrix, $\mathbb{M}^\top$ is also an M-matrix. An interesting property of an M-matrix is that all the elements of its inverse are non negative.}, i.e. such that {\it all its eigenvalues have non-negative real parts}. Therefore some conditions on productivities $z_i$ and linkages $J_{ij}, a_{ij}$ must be fulfilled for the economy to work.
Rather interestingly, Eq.  \eqref{eq:excess_prof2} is identical, {\it mutatis mutandis}, to the equation determining the equilibrium size of species in a Generalized Lotka-Volterra model discussed in section \ref{sec:GLV}.\footnote{There is a long tradition of using the {\it two-specie} classical Lotka-Volterra equation in economics, initiated by R. Goodwin, see e.g. \cite{flaschel2010classical}. However the dynamical version of the {\it multi-specie} Lotka-Volterra model has not been studied in the context of firm growth, and might be a fruitful path to follow, using in particular the recent results of Refs. \cite{herskovic2020firm, mazzarisi2024beyond}.} 

Expanding Eq.  \eqref{eq:excess_prof2} to first order in $\zeta$, one notices that since $\sum_{j > 0} a_{ij} < 1$, the Perron-Frobenius theorem ensures that $q \to \infty$ Cobb-Douglas networked economies (such as those considered in \cite{long1983real,acemoglu2012network}) {\it always} have a feasible equilibrium where all firms can survive, for any network and any productivities. Therefore, the type of crisis that takes place for $q < +\infty$, where bankruptcies occur, has no counterpart in a Cobb-Douglas economy.

In all other cases, when the smallest real part of the eigenvalues of matrix $\mathbb{M}$ reaches zero, the system approaches a critical state and bankruptcies must occur for the economy to remain viable. If the system self-stabilizes around such a critical point, then small productivity shocks can cause bankruptcy avalanches which are found to be scale-free when the production network is sufficiently heterogeneous \cite{moran2019may}.   

What drives the smallest eigenvalue of $\mathbb{M}$ to zero? Several types of evolutionary forces act to that effect. One is simply the creation of new firms, that connect to the pre-existing network at constant average productivity $\overline{z}$. As shown in \cite{moran2019may}, this can only {\it decrease} the smallest eigenvalue of the matrix $\mathbb{M}$ that describes the pre-existing firms. One concludes that a growing economy can only become more unstable with time, unless productivity increases.\footnote{For a related story, see the recent paper \cite{patil2025emergent} where the incipient instability generates emergent wealth inequalities.} This argument is actually closely related to May's original argument about the stability of large ecologies \cite{may1972will}. 

In fact, one can show that as the number of links to the most connected node of the network increases, the smallest eigenvalue of $\mathbb{M}$ decreases \cite{castellano2017relating}, until the instability threshold is reached. In this case, the fragility of the network comes from the most central hubs, a scenario akin to, but different from, the one of Acemoglu, Carvalho et al.~\cite{acemoglu2012network}. This effect might be amplified if firms systematically favour links toward hubs, leading to a ``scale-free'' free input-output network~\cite{atalay2011network}. Interestingly, a stability-constrained growth mechanism for networks, whereby a node is freely added to the network if it does not destabilize the system but induces rewirings in the network until stability is found again if it does, has been found to generate such self-organized scale-free networks~\cite{perotti2009emergent}.

The second evolutionary effect is the complexification of the production process, i.e., technology progress means that a wider array of products are needed as inputs. If the average productivity $\overline{z}$ remains the same while the average connectivity of the network increases, the system eventually reaches the instability point. Hence, again, productivity must increase at some minimum rate for the economy to remain stable. But since increasing productivity is costly, one can postulate that the average productivity $\overline{z}$ will tend to hover around the minimal viable threshold, and sometimes lagging behind, leading to occasional endogenous crises. Similarly, firms tend to optimize their portfolios of suppliers, thereby reducing their redundancy but, by the same token, reducing the effective substitution effects captured by the CES parameter $q$. As $q$ decreases, the economy will again inevitably become unstable. 

Finally, competitive equilibrium corresponds to zero profit. Now, in more realistic situations, firms attempt to realize positive profits and distribute dividends. In the present framework, this amounts (for $q=0$) to a reduction of effective productivity $z_i \to z_i(1-\varphi_i)$, where $\varphi_i$ is the fraction of the total sales firm $i$ targets as profit. As firms attempt to increase their profits, the average effective productivity goes down, until the marginal stability point is reached and a crisis ensues. After the crisis, economic actors revert to more reasonable levels of markups, which makes the economy viable again --- until the next crisis: we are again back to a Minsky cycle of sorts.   

One may therefore conjecture that evolutionary and behavioural forces repeatedly drive the economy close to marginal stability, as argued by Bak et al. \cite{bak1993aggregate, scheinkman1994self} but here within the context of a more traditional economic equilibrium model. It is now the very structure of the supply network and the non-substitutability of input goods that determines the proximity of the critical point.

\subsection{The ecology of production networks (dynamics)} \label{sec:firm_ecology_2}

Still, the above story is all within the realm of economic equilibrium, where profits are maximized, markets clear and all produced goods are consumed. As exogenous shocks buffet the economy, a new equilibrium is assumed to be almost immediately reached, corresponding to the new values of productivities. As already discussed above, this is a heroic assumption that disregards the dynamics of e.g. inventories, stock-outs and bankruptcies. At the very least one should explain the dynamical process through which equilibrium is reached, and how long one should wait before it is reached -- if it is ever reached (on this point, see the insightful introductory chapter of Ref. \cite{fisher1989disequilibrium}).\footnote{Although it lays out a research program which is in my view of fundamental importance, Franklin Fisher's 1982 book ``Disequilibrium foundations of equilibrium economics" \cite{fisher1989disequilibrium} only has 882 citations to date. This shows that mainstream economics is still not really interested in these topics, as testified by ``state-of-the-art'' macroeconomics models which completely disregard out-of-equilibrium effects, see e.g.  \cite{kaplan2018monetary, baqaee2019macroeconomic, baqaee2020productivity}. For a related discussion, see \cite{farmer2024making}.}

There is however no consensus on how to model such dynamics. A simplified, linear approach was proposed by Dessertaine et al. \cite{dessertaine2022out} to model the dynamics of small deviations away from equilibrium, where supply-demand and profit imbalances are used as restoring forces towards equilibrium. The resulting equations are of the general form Eq. \eqref{eq:OU_d}, with $x_i = \{p_i, \pi_i\}$ and $N \to 2N$ and a $2N \times 2N$ stability matrix $\mathbb{K}$. Interestingly, the smallest negative eigenvalue $-\kappa^\star$ of $\mathbb{K}$ goes to zero precisely when the smallest positive eigenvalue of $\mathbb{M}$ goes to zero, i.e., when the supply network becomes critical in the sense defined in the previous subsection \cite{dessertaine2022out}. As discussed in section \ref{sec:trivial}, as the system reaches criticality, convergence towards equilibrium becomes infinitely slow and small exogenous shocks are disproportionately amplified. Hence, as for the Generalized Lotka-Volterra dynamics, the system becomes infinitely {\it fragile} as it reaches the critical point, just before the network has to rearrange. In the present context, it means that some firms have to disappear and/or the supply network has to rewire for the economy to remain functional.\footnote{On the slow, complex dynamics of rewiring, see e.g. \cite{colon2022radical}.}     

\begin{figure}[h!]
    \centering
    \includegraphics[width = 0.6\textwidth]{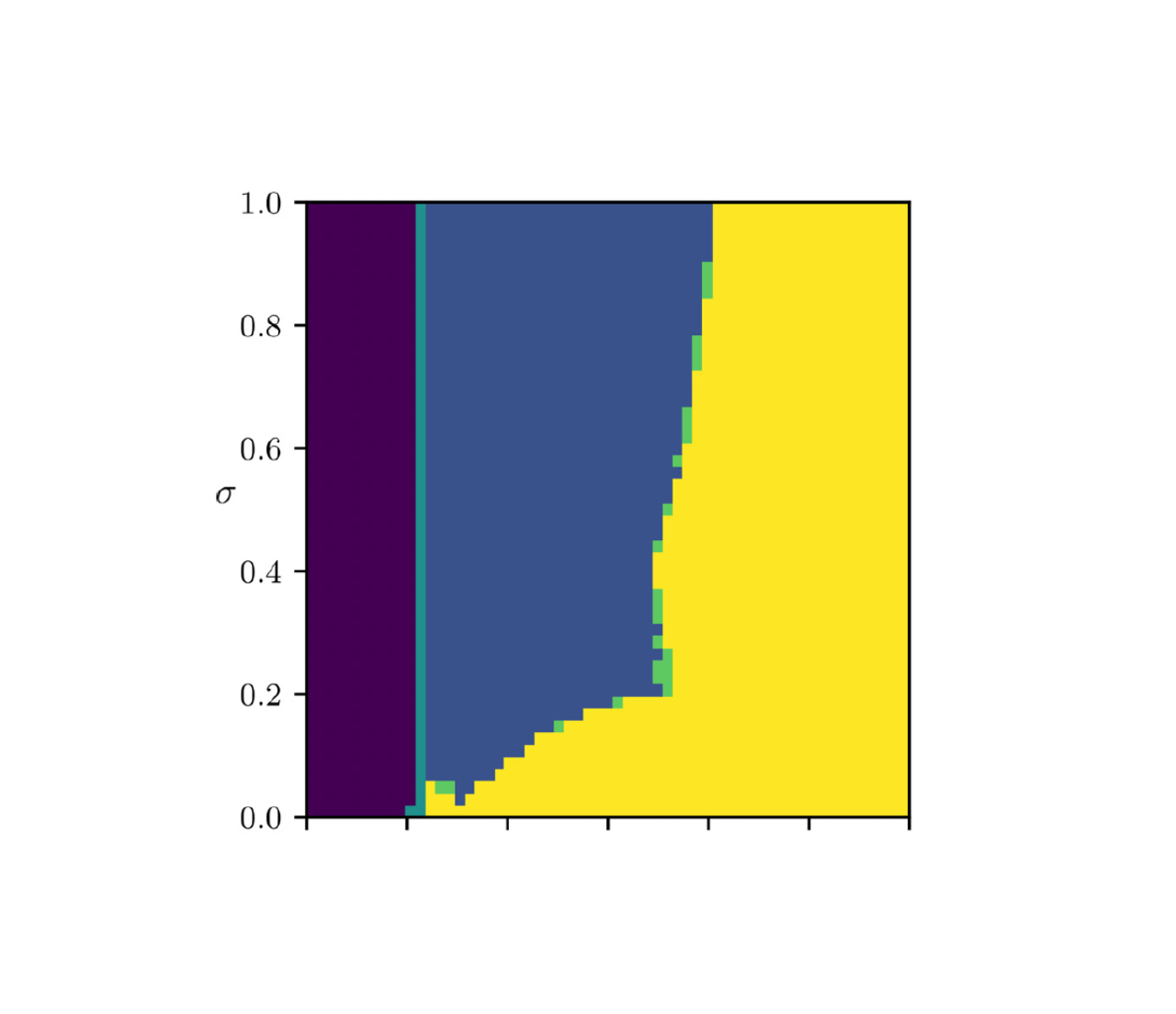}
    \caption{\small{Phase Diagram of the Agent Based firm network model of Dessertaine et al. \cite{dessertaine2022out}. $x-$axis: strength of forces counteracting supply/demand and profit imbalances; $y-$axis: goods perishability. Leftmost region (a, violet): the economy collapses; middle region (b, blue): the economy reaches equilibrium relatively quickly; righr region (c, yellow): the economy is in perpetual disequilibrium, with purely endogenous fluctuations. The green vertical sliver (d) corresponds to a deflationary equilibrium, see footnote \ref{footnote:phase_d} and \cite{dessertaine2022out}}.}
    \label{fig:phase_diagram}
\end{figure}

However, the above ``naive'', linear implementation of the out-of-equilibrium dynamics is theoretically inconsistent, as some incontrovertible constraints are violated (for example, consumption cannot exceed amount of available goods). The fully consistent, Agent Based description proposed in \cite{dessertaine2022out} goes beyond the scope of the present paper. The results are summarized in the form of the phase diagram shown in Fig. \ref{fig:phase_diagram}. Dessertaine et al. \cite{dessertaine2022out} find that, depending on the parameters of the system (strength of the reaction in the face of imbalances, perishability of goods,...) the economy can either (a) completely collapse; or (b) reach the competitive equilibrium after some time, that can become very large; or else (c) enter a state of perpetual disequilibrium, with purely endogenous fluctuations which can be periodic or completely chaotic, corresponding to a coordination breakdown of firms along the supply chain \cite{Sterman1989}.\footnote{\label{footnote:phase_d} The is actually another possibility (d) where some deflationary equilibrium is reached, see \cite{dessertaine2022out}.} In this last regime, economic equilibrium still exists but is linearly unstable, and is dynamically out-of-reach. This regime is interesting because it offers yet another explanation for ``small shocks, large business cycles”: the economy might be permanently in a turbulent state, far from equilibrium, without necessarily being close to a critical point.

The model of Dessertaine et al. \cite{dessertaine2022out} therefore suggests {\it two} distinct out-of-equilibrium routes to excess volatility: (i) purely endogenous cycles, resulting from over-reactions and non-linearities, or (ii) self-organized criticality, i.e. persistence and amplification of exogenous shocks, governed by the proximity of a boundary in parameter space where the competitive equilibrium becomes unstable. While scenario (i) may appear at first sight to be more generic, the economic forces discussed above could make (ii) plausible as well. Specific empirical work is needed to distinguish between these two scenarios, which lead to rather different predictions for the structure and dynamics of fluctuations. The SOC scenario suggests in particular that fluctuations should be correlated over long time scales, corresponding to the propagation of ``avalanches'' over large portions of the production network. 

\subsection{Inflation ``avalanches''}\label{sec:inflation}

In a very interesting recent paper, M. Nirei \& J. Scheinkman \cite{nirei2024repricing} have suggested that repricing events (i.e. firms increasing their price in the face of inflation) are clustered in time -- in other words that inflation occurs in waves, or avalanches. The idea that inflation dynamics is intermittent actually dates back to Rob Engle in 1982, and motivated the introduction of the famous ARCH model \cite{engle1982autoregressive},\footnote{See Eq. \eqref{eq:ARCH} below. The 1982 Engle paper has, according to Google scholar, 34,000 citations as of today!} before it was generalized and applied to volatility in financial markets \cite{engle2001garch}, see section \ref{sec:arch}.

Nirei \& Scheinkman develop a rather sophisticated equilibrium model to describe the repricing strategy of profit-optimizing firms. Here we simplify the model but keep its essential ingredients. We assume that the inflation rate index is $I(t) > 0$. In a mean-field spirit, we do not model the details of the client-supplier network but assume that firms are only sensitive to the overall inflation index. The {\it real} (i.e. deflated) log-price $p_i$ of firm $i$ then evolves, in the absence of repricing, as 
\[
\frac{{\rm d}p_i}{{\rm d}t} = - I(t).
\]
At one point, firms have to reprice up otherwise production costs will exceed their expected income. We assume that firm $i$ chooses to do so at a Poisson rate $\gamma$ if $p_- < p_i < p_+$, in which case the new price is $p_i = p_+ > 0$. But if $p_i$ reaches $p_-$, expected future losses are deemed too high and the firm reprices immediately at $p_+$. The evolution of the distribution of prices $P(p,t)$ is then given by (compare with Eq. \eqref{eq:slope})
\begin{equation}\label{eq:FP_inflation}
    \frac{\partial P(p,t)}{\partial t} =  I(t) \frac{\partial P(p,t)}{\partial p} - \gamma P(p,t), \qquad p \in [p_-,p_+]
\end{equation}
with an equation describing the ``reinjection'' process at $p_+$, 
\begin{equation}\label{eq:flux}
    I(t) P(p_+,t) = \gamma + I(t) P(p_-,t),
\end{equation}
which expresses that all firms that reprice, either spontaneously or constrained to do so, contribute to the probability flux at $p_+$, equal to $I(t) P(p_+,t)$.

Now, the crux of the matter is that a firm that reprices up contributes to the average inflation seen by its customers, and therefore pushes other firms towards repricing. This is the basic mechanism for repricing ``avalanches''. Let us thus write, again in a mean-field spirit, the effect of repricing on inflation as 
\begin{equation}
    I(t) = I_0 + J \Big[\gamma (p_+ - \overline{p}(t)) + I(t) P(p_-,t) (p_+ - p_-)\Big], \qquad \overline{p}(t) = \int_{p_-}^{p_+} {\rm d}p \, p P(p,t).
\end{equation}
where $I_0$ is the primary good inflation (say energy) and $J$ measures an average client-supplier coupling strength between firms in the production network.\footnote{Parameter $J$ is the direct analogue of parameter $\theta$ defined in Refs. \cite{nirei2024repricing, leal2021repricing}.} The two terms inside the square brackets correspond to spontaneous repricing (at rate $\gamma$) and forced repricing (firms with $p_i \approx p_-$).

We now look for a stationary state for the system, i.e. $P(p,t)=P_{\rm{st.}}(p)$ and $I(t)=I_{\rm{st.}}$. From Eq. \eqref{eq:FP_inflation} one derives
\[
P_{\rm{st.}}(p) = Z^{-1} \exp\left[\frac{\gamma p}{I_{\rm{st.}}}\right], \qquad Z = \frac{I_{\rm{st.}}}{\gamma} \left( \exp\left[\frac{\gamma p_+}{I_{\rm{st.}}}\right] - \exp\left[\frac{\gamma p_-}{I_{\rm{st.}}}\right]\right)
\]
where $Z$ is such that $\int_{p_-}^{p_+} {\rm d}p \,P_{\rm{st.}}(p)=1$. One can check that Eq. \eqref{eq:flux} is automatically satisfied, as it should be. In the limit $\gamma (p_+-p_-) \ll I_{\rm{st.}}$, the average inflation rate is then given by 
\begin{equation} \label{eq:inflation}
    I_{\rm{st.}} \approx  \frac{I_0}{1 - J}, \qquad (J < 1).
\end{equation}

Now, what happens when a single firm with price close to $p_-$ reprices? By doing so, it increases inflation index $I$ by an amount $J (p_+-p_-)$, such that other firms close to $p_-$ tip over and also reprice. The branching ratio $R_0$ (average number of firms tipping over) is thus given by
\begin{equation} \label{eq:R0_inflation}
    R_0 = \int_{p_-}^{p_- + J (p_+-p_-)} {\rm d}p \, P_{\rm{st.}}(p) \approx J
\end{equation}
where the last equality holds in the same limit $\gamma (p_+-p_-) \ll I_{\rm{st.}}$ considered above. 

Hence, the above repricing model becomes exactly equivalent to the branching process described in section \ref{sec:branching}. The model therefore becomes critical when $J \to 1$, as repricing avalanches of diverging size appear, leading to runaway inflation, see Eq. \eqref{eq:inflation}. When $J \to 1$, the distribution of the size $S$ of repricing avalanches is given by Eq. \eqref{eq:avalanches}, and more generally by a ``Generalized Poisson Distribution'', see \cite{nirei2024repricing}. Quite remarkably, a calibration of the model on empirical data suggests that $J$ is quite close to unity ($J \sim 0.88$ -- $0.95$ for a variety of countries) \cite{leal2021repricing}, suggesting again that the economy is in a strongly coupled regime, not far from a critical point. Interestingly, inflation volatility is also predicted to increase with inflation itself (as $I_{\rm{st.}}^{3/2}$), as empirically observed \cite{nirei2024repricing}.

Quite remarkably, exactly the same mechanism was proposed in \cite{gualdi2015endogenous} to explain the endogenous bankruptcy waves observed in the stylized macroeconomic Agent Based model ``Mark0'' proposed in \cite{gualdi2015tipping}. In that model, firms default when their debt to sales ratio exceeds a certain threshold, while new born firms appear with a positive initial balance sheet at some Poisson rate. Much like for the above inflation story, the default of a single firm deteriorates the economic environment of other firms, pushing some of them beyond their bankruptcy threshold. The mathematical encryption of this mechanism is very similar to Eq. \eqref{eq:FP_inflation}, see \cite{gualdi2015endogenous}, and is related to the propagation of neuronal activity in the brain \cite{de2006self, lombardi2021long}, which is also argued to be close to critical \cite{chialvo2010emergent, kinouchi2020mechanisms}, and for which the regime $J > 1$ corresponds to epilepsy crises. 

In Ref. \cite{gualdi2015endogenous}, special attention was devoted to the case $J > 1$, where no stationary state can be reached. Instead, an oscillating ``business cycle'' dynamics appear, where the feedback mechanism described above leads to the synchronisation of bankruptcies and spikes of unemployment, see also \cite{sharma2021good} for similar effects. One can expect similar inflation cycles to set in for $J > 1$ within the context of the Nirei-Scheinkman model, although the introduction of monetary policy would probably significantly affect the results. Still, the question of why $J$ appears to be empirically so close to unity (i.e. why is the system close to critical) is an open question that certainly requires more thought. In that respect, the analogy between economic activity and the critical brain activity, with its occasional seizures \cite{de2006self, chialvo2010emergent, osorio2010epileptic, kinouchi2020mechanisms}, is quite enticing -- after all, economic agents are, like neurons, processing information, computing prices and productions, and sending signals to other agents.

\section{Self-Organized Criticality in Finance}\label{sec:SOC_F}

Financial price series exhibit several ``anomalous'' statistical properties reminiscent of critical systems. For example, as it is well known, the probability distribution function of daily returns $r$ has fat, power-law tails both for positive and negative returns, decaying as $|r|^{-1-\alpha}$ with $\alpha$ in the range $(3,5)$ for almost all traded instruments \cite{cont2001empirical, bouchaud2001power,bouchaud2003theory,gabaix2009power} (including rather exotic ones like cryptocurrencies, Credit Default Swaps or implied volatility, see \cite{bouchaud2011endogenous, bouchaud2017have}). 

The second intriguing feature is {\it long memory}: realized volatility, market activity (measured as transaction volumes, number of price changes, or order book events), sign of market orders, all show a time auto-correlation function that can be fitted by a slow power-law. In the case of market activity, one can measure such long memory effects from minutes to years. This is captured in different ways by e.g. rough/multifractal volatility models \cite{muzy2000modelling, muzy2013random, bayer2023rough} or by Hawkes processes \cite{hardiman2014branching,bacry2015hawkes} which all rely on scale-free, slowly decaying kernels that propagate the impact of shocks forward in time. As discussed in section \ref{sec:simple}, the appearance of slow time scales (here much slower than the time between two transactions) is often a signature of critical behaviour.     

Perhaps the most striking observation suggesting some form of {\it market fragility} is the fact that most major price jumps do not seem to be related to anything particular happening in the world, at least seen from major news feeds like Bloomberg -- see e.g. \cite{cutler1988moves, fair2002events, joulin2008, marcaccioli2022exogenous, aubrun2024riding}.\footnote{\label{footnote:kyle} Note that private information should not make prices jump. Traders who want to benefit from private information should execute their orders in a stealthy manner, as clearly illustrated by Kyle's model \cite{kyle1985continuous}. By making the price jump, they would immediately lose their informational advantage.} Of course, some news does make prices jump, but most of the time jumps seem to be a result of endogenous feedback loops. Seemingly innocuous shocks get amplified, resulting in micro-, meso- or even macro-liquidity crises (see section \ref{sec:liquidity} below). Hence, markets appear to operate ``at the edge of chaos'', which would explain why they tend to overreact to small perturbations. The precise mechanism(s) driving markets close to a critical point are however not known -- or rather several such mechanisms have been proposed (as reviewed below), but none of them have been accepted on the basis of clear, falsifiable predictions.   

There are actually other, non-critical explanations for power-laws and long memory. For example, Gabaix et al. \cite{gabaix2003theory} argue that the power-law tail of the return distribution is inherited from the power-law distribution of assets under management.\footnote{The fact that the distribution of assets is a power-law could itself be a signature of criticality. However, simple plausible models based on random multiplicative growth can easily explain such a behaviour -- see the discussion in  \cite{bouchaud2000wealth} and \cite{schwarzkopf2010empirical}.} Large investment firms trade large quantities and  should induce large price moves. This idea, however, does not seem to fly: large order sizes are sliced and diced in such a way to be commensurate to the trading volume (see footnote \ref{footnote:kyle}). Hence those large orders do not create large return, but rather long-range correlations in the sign of market orders, see \cite{lillo2005theory,bouchaud2009markets, bouchaud2018trades, sato2023inferring}. Furthermore, large price moves seem to be related to sudden liquidity shortages, not to outsized incoming volumes, see \cite{farmer2004origin, gillemot2006there, joulin2008, fosset2020endogenous}.    

Long memory effects, on the other hand, have been attributed to the existence of several natural time scales in human activity, more or less distributed on a log-scale (hours, days, weeks, months, quarters, years...) \cite{bouchaud2000wealth, bochud2007optimal}. If different market participants are somehow tuned on these human frequencies, multi-time scales should naturally transpire in market activity and volatility \cite{zumbach2001heterogeneous}. Although such effects might indeed contribute to long memory (there is, for example, a clear periodicity in activity/volatility due to quarterly earning announcements), excess volatility and volatility clustering seem to be two sides of the same coin, related to self-excitation effects inherent to financial markets, as we discuss now.

\subsection{ARCH \& Hawkes processes}\label{sec:arch}

That endogenous feedback loops may be responsible for the excess volatility of financial markets has been suggested many times in the past, see e.g. \cite{bouchaud2011endogenous, fosset2020endogenous, Wehrli2022}. In fact, the classic ARCH framework mentioned in section \ref{sec:inflation} is arguably the simplest way to describe such a feedback loop from past realized volatility to future volatility. The model posits that daily returns can be written as $r_t = \sigma_t \xi_t$ where $\xi_t$ is an IID $N(0,1)$ Gaussian noise and $\sigma_t$ the time dependent volatility. One then postulates the following feedback equation:
\begin{equation}\label{eq:ARCH}
    \sigma_t^2 = \sigma_0^2 + \sum_{t'< t} \Phi(t-t') \,r_{t'}^2,
\end{equation}
where $\Phi(.)$ is a kernel describing how the perceived volatility $r_{t'}^2$ at time $t'<t$ impacts volatility at time $t$. Averaging Eq. \eqref{eq:ARCH} over time and assuming stationarity, one finds that the average square volatility $\sigma_\infty^2 = \mathbb{E}_t[\sigma_t^2]$ is given by 
\begin{equation}
    \sigma_\infty^2 = \frac{\sigma_0^2}{1 - g}; \qquad g:=\sum_{\tau=1}^\infty \Phi(\tau) < 1,
\end{equation}
showing that feedback, measured by parameter $g$, leads to excess volatility. When $g \nearrow 1$, self-referential effects become so strong that volatility explodes (compare with Eq. \eqref{eq:inflation}). $g=1$ again appears as a critical point, in fact accompanied by slowing down, as in section \ref{sec:simple}. More precisely, the autocorrelation function of $\sigma_t^2$ decays over a time scale much longer than the decay time of the kernel $\Phi(\tau)$ when $g \nearrow 1$.

Hawkes processes provide an alternative description for self-exciting feedback loops. A Hawkes process describes Poisson events whose intensity increases with the past activity of the process itself (for a review in the context of financial markets, see \cite{bacry2015hawkes}). Such an increase is mediated by a kernel $\Phi(\tau)$, as for ARCH processes, but now in continuous time. When $g := \int {\rm d}\tau \, \Phi(\tau)$ becomes larger than unity, the intensity of the process diverges. When $g$ is precisely equal to $1$ the process is critical and only exists for long memory (power-law) kernels \cite{bremaud2001hawkes}.

Both ARCH and Hawkes models have been calibrated on data. In all cases, one finds very high values $g \gtrsim 0.8$ for the integrated feedback parameter (see e.g. \cite{chicheportiche2014fine, hardiman2013critical, filimonov2014quantification, hardiman2014branching, blanc2017quadratic, wheatley2019endo, Wehrli2022}), with a power-law decaying kernel $\Phi(\tau)$. This is in line with Shiller's excess volatility puzzle \cite{shiller1981stock,shiller1987volatility, leroy1981present}: volatility in financial markets is at least five times larger than what it ``should'' be in the absence of feedback -- and as noted in the previous section, many jumps appear without obvious causes. 

But yet again, one is entitled to ask: why is the feedback kernel a power-law and the feedback parameter so high? What are the driving forces explaining why the system appears to be not far from criticality, as we saw in the case of inflation as well?

\subsection{Critical liquidity provision}\label{sec:liquidity}

One plausible explanation for the fragility of financial markets is that at the very heart of all markets lies a fundamental competitive tension: buyers want to buy low and sellers want to sell high. The role of {\it market makers} (MM), also known as liquidity providers, is to resolve (part of) this tension. MM buy from sellers and sell later to buyers (or vice versa), allowing markets to solve the immediacy problem and ensure, most of the time, ``fair and orderly trading''. MM make a living from the bid-ask spread $s_t$ -- buying slightly below current mid-price $p_t$ and selling slightly above $p_t$, while trying to keep their inventory as close to zero as possible. 

Market-makers however bear the risk of adverse selection, which results from the fact they must post binding quotes (bid or ask prices), which can be ``picked off” by more informed traders who see an opportunity to buy low or to sell high. In other words, if the MM sells (/buys) at a price $p_t \pm s_t/2$ that turns out be lower (higher) than the future price at which they will need to unwind their position, they will suffer a loss that can be very large. More formally, let us define the average adverse price move as
\begin{equation}
    \Delta := \mathbb{E}\left[\epsilon_t \cdot (p_\infty - p_t)\right],
\end{equation}
where $\epsilon_t$ is the sign of the incoming trade and $p_\infty$ is the price in a sufficiently distant future (for more precise statements, see \cite{wyart2008relation} and \cite{bouchaud2018trades}, ch. 17).  
Market-making is thus only profitable if the bid-ask spread exceeds, on average, adverse selection, i.e. if $\mathbb{E}[s_t]/2 > \Delta$. 

Unfortunately for market-makers, the distribution of price changes after a trade is very broad, with a heavy tail in the direction of the trade. Whereas most trades contain relatively little information and are therefore innocuous for MM, some rare trades
are followed by extreme adverse moves, wiping out the small profits accumulated on ``normal'' transactions. In other words, the distribution of MM profits is strongly negatively skewed. 

Because liquidity providers compete, the spread is compressed to values close to break even, i.e. $s \approx 2\Delta$. But since the risk of large losses is huge, MM are quick to increase the spread and reduce the amount of liquidity that they offer for purchase or sale when volatility ticks up.

This simple argument illustrates why liquidity is fragile and can disappear quickly.\footnote{As the saying goes: {\it Liquidity is a coward. It is never there when it is needed.}} In fact, the classic Glosten-Milgrom model \cite{glosten1985bid} predicts that high volatilities can lead to liquidity crises, in the sense that no value of the spread can allow MM to break even. Hence, there is a built-in destabilising feedback loop, intrinsic to the way markets operate: more volatility leads to higher spreads and less liquidity, which itself leads to higher volatilities. Such a mechanism is discussed and empirically validated in \cite{fosset2020endogenous, dall2019does, fosset2022non}, and sounds like a plausible explanation for the large fraction of endogenous price jumps mentioned above \cite{joulin2008, marcaccioli2022exogenous, aubrun2024riding}.

Markets are thus in a subtle equilibrium between liquidity takers (the excitatory force) and liquidity providers (the inhibitory force) \cite{bouchaud2003fluctuations}, as for the balancing stick problem of section \ref{sec:stabilizing}, or for heart beats or neuronal activity \cite{cabrera2004human, cabrera2012stick, patzelt2011criticality, lombardi2017balance}. In such a story, criticality in financial markets is enforced by competition that drives the system in the vicinity of the break-even point $s = 2\Delta$, such that neither liquidity takers nor liquidity providers are systematically favoured \cite{wyart2008relation,bouchaud2018trades}. Much like in section \ref{sec:timeliness}, efficiency (i.e. small spreads offered to liquidity takers) comes with instability. It would be rewarding to come up with a stylized model, in the spirit of section \ref{sec:timeliness}, that would neatly encapsulate the above scenario and allow one to compute, for example, the distribution of the size of endogenous jumps, i.e. when liquidity spontaneously disappears through feedback effects. 

\subsection{Competitive market ecology}\label{sec:ecology}

Another path to understand the apparent criticality of financial markets is to think in terms of an ecological, competitive equilibrium between different investment strategies \cite{farmer2002market}. The paper of Lux \& Marchesi \cite{lux1999scaling} is the first to explicitly exploit this idea and insist on the (self-organized) criticality of the resulting artificial market, see also \cite{scholl2021market} for closely related ideas. 

Following C. Chiarella \cite{chiarella1992dynamics}, investors are split into three broad classes: value investors (arbitraging the difference between market price and their estimate of value), trend followers (buying when the price went up, and selling when it went down) and noise traders (with essentially random strategies). Lux \& Marchesi add the possibility for investors to switch between trend following and value depending on the relative pay-off of the two strategies (see also \cite{palmer1994artificial, brock1998heterogeneous}). Quite interestingly, these ingredients appear to be enough to generate fat-tailed returns and long-range volatility correlations, which come from the fact that agents can stick to one kind of strategy for very long times -- see \cite{giardina2003bubbles} for more details on this point. The co-existence of trend followers and value traders is further documented empirically in \cite{bouchaud2017black, majewski2020, schmitt2021trend, lux2021can} and refs therein. 

There are several variations on the same theme. For example, the ``Minority Game'' is a multi-agent stylized model of market ecology amenable to analytic calculations \cite{challet2004minority}, which reveals a {\it genuine phase transition} between ``no-arbitrage'' and exploitable regularities. As the number of agents $M$ playing the game increases, profit opportunities wither and actually vanish for a critical value $M_c$, beyond which the average profit of agents becomes negative due to transaction costs. 

The attractive nature of the critical point is then quite clear \cite{challet2003criticality, giardina2003bubbles, alfi2009self} and somewhat akin to the classic Grossman-Stiglitz paradox \cite{grossman1980impossibility}: either $M < M_c$ and new investors are lured in by profit opportunities, such that $M \nearrow \,$; or 
$M > M_c$, current investors then make losses and progressively leave the market, until $M \approx M_c$. As markets become more complex and unpredictable, they also become more fragile, as discussed in section \ref{sec:stabilizing}. 

This generic path to criticality is actually quite enticing and lends credence to the SOC scenario for financial markets, see \cite{alfi2009self} for a review, and \cite{farmer2002market,  amir2005market, caccioli2009eroding, patzelt2011criticality, marsili2014complexity} for related ideas, in particular the difficulty for prices to reach equilibrium when incentives provided by mispricing tend to zero \cite{grossman1980impossibility, cherkashin2009reality, scholl2021market}. 

Note finally that a Generalized Lotka-Volterra description of the financial market ecology was derived long ago by Doyne Farmer \cite{farmer2002market}, along the lines of Eq. \eqref{eq:GLV}. As market participants trade and impact prices, other participants either benefit or suffer from those price changes. This could provide another justification of market criticality: investment strategies feed onto one another in mutually beneficial or predator-prey relationships, co-evolving into a marginally stable equilibrium, as in section \ref{sec:GLV}.     

\subsection{Contagion \& financial stability}\label{sec:financial_contagion}

Finally, let us mention that contagion mechanisms are evidently at play in financial markets, much like in the banking sector \cite{haldane2011systemic,gai2010contagion,caccioli2018network}. As we just mentioned, the trades of one investor impacts the holdings of all other market participants. When two investors (say Alice \& Bob) have {\it overlapping portfolios}, i.e. similar positions in the market, the  deleveraging of Alice's portfolio will lead to a depreciation of the marked-to-market value of Bob's portfolio \cite{caccioli2012impact,cont2013running, caccioli2015overlapping}. If the drop of value is too large, Bob may be tempted (or compelled by his broker) to deleverage his own portfolio, lowering further the value of Alice's holdings, and perhaps of other investors as well. Such a contagion mechanism can set off ``deleveraging spirals'', i.e. avalanches of trades in the same direction, collectively detrimental to all investors holding similar positions. Such a mechanism has been argued to be at the origin of the infamous ``quant crunch'' of 2007 \cite{khandani2007happened}.

The basic model for such a phenomenon is again the branching process of section \ref{sec:branching}. The contagion parameter $R_0$ is then a measure of the similarity between portfolios and of the sensitivity of funds to losses. If leverage constraints are binding, for example, $R_0$ will be higher. It is not clear, however, why in this story $R_0$ should be particularly close to the critical value $R_0=1$. But one could easily argue with Minsky that as complacency increases, portfolio risk also increases making them more exposed to downside events -- thereby realizing the growing sand pile scenario of section \ref{sec:sweeping}.    

\section{Conclusion. Efficiency vs. Resilience}\label{sec:conclusion}

Elucidating the very mechanisms leading to instabilities, failures and system-wide crises in socio-economic systems is crucial to finding remedies, mitigation measures and proposing adequate regulations. Indeed, as we have argued throughout this paper and in Ref. \cite{moran2023temporal}, the quest for ``efficiency'' and ``optimality'' and the necessity of {\it resilience}, i.e. of robustness against small perturbations, may be mutually incompatible, a point forcefully made in \cite{carlson1999highly}. 

The Self-Organized Criticality scenario  goes even further, as it suggests that optimisation necessarily leads in many complex systems to a marginally stable point that is particularly {\it fragile} -- i.e. at the edge of the proverbial cliff. We have seen for example how ``just-in-time'' policies, low inventories or over-optimized timetables necessarily push the system towards a functioning point where disruption ``avalanches'' of all sizes can appear (sections \ref{sec:timeliness}, \ref{sec:firm_ecology_1}). We have argued that liquidity in financial markets is fragile because competition between market makers drive their profits close to zero, making them hyper-sensitive to volatility blips (section \ref{sec:liquidity}). We have discussed how competitive ecologies (of living systems, of firms, of economic agents, of investment strategies) tend to co-evolve towards marginally stable, fragile equilibrium points where just enough species have to disappear for others to survive (sections \ref{sec:GLV}, \ref{sec:firm_ecology_1}, \ref{sec:firm_ecology_2}, \ref{sec:ecology}). Finally, we have emphasized how stabilising unstable systems may be inherently difficult, and may generate anomalously large fluctuations when stabilisation occasionally falters (sections \ref{sec:stabilizing}, \ref{sec:liquidity}). More generally, systems driven by antagonist forces (excitatory and inhibitory) tend to operate close to criticality. The point is that fragility and flexibility/adaptability are often two sides of the same coin.

Several plausible mechanisms could thus bring sufficiently often financial markets and economies as a whole close to, or even through a critical point where the system loses stability. At this juncture, however, we have many more narratives than hard empirical evidence for such an interpretation of the pervasive ``excess volatility'', or ``small shocks, large business cycle'' effects routinely observed in these systems. A renewed data-driven effort, like in \cite{leal2021repricing}, is certainly needed to convince the economics community of the relevance of {\it fragility} to understand excess fluctuations, without having to invoke major exogenous shocks (that are often difficult to identify and  substantiate).     

Similarly, we also sometimes lack clear mechanisms explaining why the system is attracted to the critical point, for example in the case of inflation avalanches, see section \ref{sec:inflation} \cite{nirei2024repricing}.
We have actually mentioned several times that excess fluctuations could be due to other mechanisms, {\it not} related to the immediate proximity of a critical point. The system might be such that its dynamics cycles through an instability, like a slowly growing sand pile that periodically collapses or, in a financial context, a growing complacency towards risk that leads to market crashes -- the famous Minsky cycle \cite{minsky}. Another possibility is that the system is linearly unstable but non-linearly stable, chaotically evolving in a region of parameter space and generating {\it purely endogenous} fluctuations, as we discussed in the context of firm networks in section \ref{sec:firm_ecology_2}. Yet another path is provided by the role of noise in bi-stable systems \cite{di2016self,  harras2012noise}. 

In any case, the main policy consequence of fragility in socio-economic systems is that any welfare function that system operators, policy makers of regulators seek to optimize should contain a measure of the robustness of the solution to small perturbations, or to the uncertainty about parameters value.  Adding such a resilience penalty will for sure increase costs and degrade strict economic performance, but will keep the solution at a safe distance away from the cliff edge. As argued by Taleb \cite{taleb}, and also using a different language in Ref. \cite{Hynes2022}, good policies should ideally lead to ``anti-fragile'' systems, i.e., systems that spontaneously improve when buffeted by large shocks.

\subsection*{Acknowledgements} I wish to thank Doyne Farmer for asking me to write this piece and allowing me to put my ideas together on this exciting topic. His comments on my initial draft have been very useful. My understanding of the subject owes a lot to many conversations with him over the years, and also with many friends and collaborators: F. Aguirre-Lopez, C. Aubrun, E. Bacry, M. Benzaquen, G. Biroli, J. Bonart, G. Bunin, F. Caccioli, D. Challet, C. Colon, Th. Dessertaine, J. Donier, A. Fosset, X. Gabaix, J. Garnier-Brun, J. Gatheral, I. Giardina, M. Gould, S. Gualdi, S. Hardiman, K. Kanazawa, A. Kirman, P. Le Doussal, A. Majewski, R. Marcaccioli, M. Marsili, I. Mastromatteo, M. Mézard, J. Moran, R. Morel, J.-F. Muzy, D. Panja, N. Patil, F. Patzelt, F. Pijpers, M. Potters, M. Rosenbaum, J. Scheinkman, A. Secchi, J. Sethna, M. Smerlak, D. Sornette, N. Taleb, M. Tarzia, U. Weitzel, M. Wyart \& F. Zamponi.

\bibliographystyle{unsrt}
\bibliography{references}  

\providecommand{\noopsort}[1]{}\providecommand{\singleletter}[1]{#1}%
\begin{thebibliography}{100}

\bibitem{bak1993aggregate}
Per Bak, Kan Chen, Jos{\'e} Scheinkman, and Michael Woodford.
\newblock Aggregate fluctuations from independent sectoral shocks: self-organized criticality in a model of production and inventory dynamics.
\newblock {\em Ricerche economiche}, 47(1):3--30, 1993.

\bibitem{bak2013nature}
Per Bak.
\newblock {\em How nature works: the science of self-organized criticality}.
\newblock Springer Science \& Business Media, 2013.

\bibitem{bak1987self}
Per Bak, Chao Tang, and Kurt Wiesenfeld.
\newblock Self-organized criticality: An explanation of the 1/f noise.
\newblock {\em Physical review letters}, 59(4):381, 1987.

\bibitem{sornette1989self}
Anne Sornette and Didier Sornette.
\newblock Self-organized criticality and earthquakes.
\newblock {\em Europhysics Letters}, 9(3):197, 1989.

\bibitem{frisch1995turbulence}
Uriel Frisch.
\newblock {\em Turbulence: the legacy of AN Kolmogorov}.
\newblock Cambridge university press, 1995.

\bibitem{sethna2001crackling}
James~P Sethna, Karin~A Dahmen, and Christopher~R Myers.
\newblock Crackling noise.
\newblock {\em Nature}, 410(6825):242--250, 2001.

\bibitem{sachs2012black}
MK~Sachs, MR~Yoder, DL~Turcotte, JB~Rundle, and BD~Malamud.
\newblock Black swans, power laws, and dragon-kings: Earthquakes, volcanic eruptions, landslides, wildfires, floods, and soc models.
\newblock {\em The European Physical Journal Special Topics}, 205:167--182, 2012.

\bibitem{watkins201625}
Nicholas~W Watkins, Gunnar Pruessner, Sandra~C Chapman, Norma~B Crosby, and Henrik~J Jensen.
\newblock 25 years of self-organized criticality: concepts and controversies.
\newblock {\em Space Science Reviews}, 198:3--44, 2016.

\bibitem{may1972will}
Robert~M May.
\newblock Will a large complex system be stable?
\newblock {\em Nature}, 238(5364):413--414, 1972.

\bibitem{bak1997mass}
Per Bak and Maya Paczuski.
\newblock Mass extinctions vs. uniformitarianism in biological evolution.
\newblock In {\em Physics of Biological Systems: From Molecules to Species}, pages 341--356. Springer, 1997.

\bibitem{de2006self}
Lucilla De~Arcangelis, Carla Perrone-Capano, and Hans~J Herrmann.
\newblock Self-organized criticality model for brain plasticity.
\newblock {\em Physical review letters}, 96(2):028107, 2006.

\bibitem{chialvo2010emergent}
Dante~R Chialvo.
\newblock Emergent complex neural dynamics.
\newblock {\em Nature physics}, 6(10):744--750, 2010.

\bibitem{osorio2010epileptic}
Ivan Osorio, Mark~G Frei, Didier Sornette, John Milton, and Ying-Cheng Lai.
\newblock Epileptic seizures: quakes of the brain?
\newblock {\em Physical Review E}, 82(2):021919, 2010.

\bibitem{kinouchi2020mechanisms}
Osame Kinouchi, Renata Pazzini, and Mauro Copelli.
\newblock Mechanisms of self-organized quasicriticality in neuronal network models.
\newblock {\em Frontiers in Physics}, 8:583213, 2020.

\bibitem{bialek2014social}
William Bialek, Andrea Cavagna, Irene Giardina, Thierry Mora, Oliver Pohl, Edmondo Silvestri, Massimiliano Viale, and Aleksandra~M Walczak.
\newblock Social interactions dominate speed control in poising natural flocks near criticality.
\newblock {\em Proceedings of the National Academy of Sciences}, 111(20):7212--7217, 2014.

\bibitem{dekker2021cascading}
Mark~M Dekker and Debabrata Panja.
\newblock Cascading dominates large-scale disruptions in transport over complex networks.
\newblock {\em PLoS One}, 16(1):e0246077, 2021.

\bibitem{laval2023self}
Jorge~A Laval.
\newblock Self-organized criticality of traffic flow: Implications for congestion management technologies.
\newblock {\em Transportation Research Part C: Emerging Technologies}, 149:104056, 2023.

\bibitem{Moran2024}
Jos\'e Moran, Matthijs Romeijnders, Pierre Le~Doussal, Frank~P. Pijpers, Utz Weitzel, Debabrata Panja, and Jean-Philippe Bouchaud.
\newblock Timeliness criticality in complex systems.
\newblock {\em Nature Physics, Advance Online publication, doi: https://doi.org/10.1038/s41567-024-02525-w}, 2024.

\bibitem{sornette2006endogenous}
Didier Sornette.
\newblock Endogenous versus exogenous origins of crises.
\newblock {\em Extreme events in nature and society}, pages 95--119, 2006.

\bibitem{cochrane1994shocks}
John~H Cochrane.
\newblock Shocks.
\newblock In {\em Carnegie-Rochester Conference series on public policy}, volume~41, pages 295--364. Elsevier, 1994.

\bibitem{scheinkman1994self}
Jose~A Scheinkman and Michael Woodford.
\newblock Self-organized criticality and economic fluctuations.
\newblock {\em The American Economic Review}, 84(2):417--421, 1994.

\bibitem{Bernanke1996}
B.~S. Bernanke, M.~Gertler, and S.~Gilchrist.
\newblock The financial accelerator and the flight to quality.
\newblock {\em The Review of Economics and Statistics}, 78:1--15, May 1996.

\bibitem{shiller1981stock}
Robert~J Shiller.
\newblock Do stock prices move too much to be justified by subsequent changes in dividends?
\newblock {\em American Economic Review}, 7:421, 1981.

\bibitem{shiller1987volatility}
Robert~J Shiller.
\newblock The volatility of stock market prices.
\newblock {\em Science}, 235:4784, 1987.

\bibitem{leroy1981present}
Stephen~F LeRoy and Richard~D Porter.
\newblock The present-value relation: Tests based on implied variance bounds.
\newblock {\em Econometrica: journal of the Econometric Society}, pages 555--574, 1981.

\bibitem{leroy2006excess}
Stephen~F LeRoy.
\newblock Excess volatility.
\newblock {\em The New Palgrave Dictionary of Economics, 2nd Edition. Palgrave Macmillan}, 13, 2006.

\bibitem{cutler1988moves}
David~M Cutler, James~M Poterba, and Lawrence~H Summers.
\newblock {\em What moves stock prices?}, volume 487.
\newblock National Bureau of Economic Research Cambridge, Massachusetts, 1988.

\bibitem{joulin2008}
Armand Joulin, Augustin Lefevre, Daniel Grunberg, and Jean-Philippe Bouchaud.
\newblock Stock price jumps: news and volume play a minor role.
\newblock {\em Wilmott Magazine}, 46, 2008.

\bibitem{marcaccioli2022exogenous}
Riccardo Marcaccioli, Jean-Philippe Bouchaud, and Michael Benzaquen.
\newblock Exogenous and endogenous price jumps belong to different dynamical classes.
\newblock {\em Journal of Statistical Mechanics: Theory and Experiment}, 2022(2):023403, 2022.

\bibitem{muzy2000modelling}
Jean-Fran{\c{c}}ois Muzy, Jean Delour, and Emmanuel Bacry.
\newblock Modelling fluctuations of financial time series: from cascade process to stochastic volatility model.
\newblock {\em The European Physical Journal B-Condensed Matter and Complex Systems}, 17:537--548, 2000.

\bibitem{cont2001empirical}
Rama Cont.
\newblock Empirical properties of asset returns: stylized facts and statistical issues.
\newblock {\em Quantitative finance}, 1(2):223, 2001.

\bibitem{bouchaud2003theory}
Jean-Philippe Bouchaud and Marc Potters.
\newblock {\em Theory of financial risks}, volume~12.
\newblock Cambridge University Press, Cambridge, 2003.

\bibitem{gabaix2009power}
Xavier Gabaix.
\newblock Power laws in economics and finance.
\newblock {\em Annu. Rev. Econ.}, 1(1):255--294, 2009.

\bibitem{parisi2007physics}
Giorgio Parisi.
\newblock Physics complexity and biology.
\newblock {\em Advances in Complex Systems}, 10(supp02):223--232, 2007.

\bibitem{PARISI1999557}
Giorgio Parisi.
\newblock Complex systems: a physicist's viewpoint.
\newblock {\em Physica A: Statistical Mechanics and its Applications}, 263(1):557--564, 1999.
\newblock Proceedings of the 20th IUPAP International Conference on Statistical Physics.

\bibitem{bouchaud2021radical}
Jean-Philippe Bouchaud.
\newblock Radical complexity.
\newblock {\em Entropy}, 23(12):1676, 2021.

\bibitem{gabaix2011granular}
Xavier Gabaix.
\newblock The granular origins of aggregate fluctuations.
\newblock {\em Econometrica}, 79(3):733--772, 2011.

\bibitem{moran2024revisiting}
Jos{\'e} Moran, Angelo Secchi, and Jean-Philippe Bouchaud.
\newblock Revisiting granular models of firm growth.
\newblock {\em arXiv preprint arXiv:2404.15226}, 2024.

\bibitem{axtell2001zipf}
Robert~L Axtell.
\newblock Zipf distribution of us firm sizes.
\newblock {\em science}, 293(5536):1818--1820, 2001.

\bibitem{gabaix2003theory}
Xavier Gabaix, Parameswaran Gopikrishnan, Vasiliki Plerou, and H~Eugene Stanley.
\newblock A theory of power-law distributions in financial market fluctuations.
\newblock {\em Nature}, 423(6937):267--270, 2003.

\bibitem{gabaix2006institutional}
Xavier Gabaix, Parameswaran Gopikrishnan, Vasiliki Plerou, and H~Eugene Stanley.
\newblock Institutional investors and stock market volatility.
\newblock {\em The Quarterly Journal of Economics}, 121(2):461--504, 2006.

\bibitem{schwarzkopf2010empirical}
Yonathan Schwarzkopf and J~Doyne Farmer.
\newblock Empirical study of the tails of mutual fund size.
\newblock {\em Physical Review E—Statistical, Nonlinear, and Soft Matter Physics}, 81(6):066113, 2010.

\bibitem{carlson1999highly}
Jean~M Carlson and John Doyle.
\newblock Highly optimized tolerance: A mechanism for power laws in designed systems.
\newblock {\em Physical Review E}, 60(2):1412, 1999.

\bibitem{di2016self}
Serena di~Santo, Raffaella Burioni, Alessandro Vezzani, and Miguel~A Munoz.
\newblock Self-organized bistability associated with first-order phase transitions.
\newblock {\em Physical review letters}, 116(24):240601, 2016.

\bibitem{harras2012noise}
Georges Harras, Claudio~J Tessone, and Didier Sornette.
\newblock Noise-induced volatility of collective dynamics.
\newblock {\em Physical Review E—Statistical, Nonlinear, and Soft Matter Physics}, 85(1):011150, 2012.

\bibitem{harris1963theory}
Theodore~Edward Harris et~al.
\newblock {\em The theory of branching processes}, volume~6.
\newblock Springer Berlin, 1963.

\bibitem{keeler1986robust}
James~D Keeler and J~Doyne Farmer.
\newblock Robust space-time intermittency and 1f noise.
\newblock {\em Physica D: Nonlinear Phenomena}, 23(1-3):413--435, 1986.

\bibitem{jensen1998self}
Henrik~Jeldtoft Jensen.
\newblock {\em Self-organized criticality: emergent complex behavior in physical and biological systems}, volume~10.
\newblock Cambridge university press, 1998.

\bibitem{sornette1994sweeping}
Didier Sornette.
\newblock Sweeping of an instability: an alternative to self-organized criticality to get powerlaws without parameter tuning.
\newblock {\em Journal de Physique I}, 4(2):209--221, 1994.

\bibitem{sornette2012dragon}
Didier Sornette and Guy Ouillon.
\newblock Dragon-kings: mechanisms, statistical methods and empirical evidence.
\newblock {\em The European Physical Journal Special Topics}, 205(1):1--26, 2012.

\bibitem{cavagna1998stationary}
Andrea Cavagna, Irene Giardina, and Giorgio Parisi.
\newblock Stationary points of the thouless-anderson-palmer free energy.
\newblock {\em Physical Review B}, 57(18):11251, 1998.

\bibitem{muller2015marginal}
Markus M{\"u}ller and Matthieu Wyart.
\newblock Marginal stability in structural, spin, and electron glasses.
\newblock {\em Annu. Rev. Condens. Matter Phys.}, 6(1):177--200, 2015.

\bibitem{Franz2017}
S.~Franz, G.~Parisi, M.~Sevelev, P.~Urbani, and F.~Zamponi.
\newblock Universality of the sat-unsat (jamming) threshold in non-convex continuous constraint satisfaction problems.
\newblock {\em SciPost Physics}, 2:019, 2017.

\bibitem{bunin2017ecological}
Guy Bunin.
\newblock Ecological communities with lotka-volterra dynamics.
\newblock {\em Physical Review E}, 95(4):042414, 2017.

\bibitem{biroli2018marginally}
Giulio Biroli, Guy Bunin, and Chiara Cammarota.
\newblock Marginally stable equilibria in critical ecosystems.
\newblock {\em New Journal of Physics}, 20(8):083051, 2018.

\bibitem{stone2018feasibility}
Lewi Stone.
\newblock The feasibility and stability of large complex biological networks: a random matrix approach.
\newblock {\em Scientific reports}, 8(1):1--12, 2018.

\bibitem{aspelmeier2019realizable}
T~Aspelmeier and M~A Moore.
\newblock Realizable solutions of the {Thouless-Anderson-Palmer} equations.
\newblock {\em Physical Review E}, 100(3):032127, 2019.

\bibitem{patil2025emergent}
Nirbhay Patil and Jean-Philippe Bouchaud.
\newblock Emergent inequalities in a primitive agent-based good-exchange model.
\newblock {\em arXiv preprint arXiv:2405.18116}, 2024.

\bibitem{PhysRevX.14.021039}
J\'er\^ome Garnier-Brun, Michael Benzaquen, and Jean-Philippe Bouchaud.
\newblock Unlearnable games and ``satisficing'' decisions: A simple model for a complex world.
\newblock {\em Phys. Rev. X}, 14:021039, Jun 2024.

\bibitem{fisher1991directed}
Daniel~S. Fisher and David~A. Huse.
\newblock Directed paths in a random potential.
\newblock {\em Phys. Rev. B}, 43:10728--10742, May 1991.

\bibitem{aspelmeier2008bond}
T~Aspelmeier.
\newblock Bond chaos in the {Sherrington--Kirkpatrick} model.
\newblock {\em Journal of Physics A: Mathematical and Theoretical}, 41(20):205005, 2008.

\bibitem{krzkakala2005disorder}
F~Krzakala and J-P Bouchaud.
\newblock Disorder chaos in spin glasses.
\newblock {\em Europhysics Letters}, 72(3):472, 2005.

\bibitem{garnier2021new}
Jerome Garnier-Brun, Michael Benzaquen, Stefano Ciliberti, and Jean-Philippe Bouchaud.
\newblock A new spin on optimal portfolios and ecological equilibria.
\newblock {\em Journal of Statistical Mechanics: Theory and Experiment}, 2021(9):093408, 2021.

\bibitem{patzelt2011criticality}
Felix Patzelt and Klaus Pawelzik.
\newblock Criticality of adaptive control dynamics.
\newblock {\em Physical Review Letters}, 107(23):238103, 2011.

\bibitem{cabrera2004human}
Juan~Luis Cabrera and John~G Milton.
\newblock Human stick balancing: tuning l{\'e}vy flights to improve balance control.
\newblock {\em Chaos: An Interdisciplinary Journal of Nonlinear Science}, 14(3):691--698, 2004.

\bibitem{cabrera2012stick}
JL~Cabrera and JG~Milton.
\newblock Stick balancing, falls and dragon-kings.
\newblock {\em The European Physical Journal Special Topics}, 205(1):231--241, 2012.

\bibitem{patzelt2013inherent}
Felix Patzelt and Klaus Pawelzik.
\newblock An inherent instability of efficient markets.
\newblock {\em Scientific reports}, 3(1):2784, 2013.

\bibitem{lombardi2017balance}
Fabrizio Lombardi, Hans~J Herrmann, and Lucilla de~Arcangelis.
\newblock Balance of excitation and inhibition determines 1/f power spectrum in neuronal networks.
\newblock {\em Chaos: An Interdisciplinary Journal of Nonlinear Science}, 27(4), 2017.

\bibitem{minsky2008stabilizing}
Hyman~P Minsky and Henry Kaufman.
\newblock {\em Stabilizing an unstable economy}, volume~1.
\newblock McGraw-Hill New York, 2008.

\bibitem{acemoglu2012network}
Daron Acemoglu, Vasco~M Carvalho, Asuman Ozdaglar, and Alireza Tahbaz-Salehi.
\newblock The network origins of aggregate fluctuations.
\newblock {\em Econometrica}, 80(5):1977--2016, 2012.

\bibitem{Colon2017}
C{\'{e}}lian Colon and Michael Ghil.
\newblock Economic networks: Heterogeneity-induced vulnerability and loss of synchronization.
\newblock {\em Chaos: An Interdisciplinary Journal of Nonlinear Science}, 27(12):126703, December 2017.

\bibitem{minsky}
Hyman~P Minsky.
\newblock {\em Can ``it'' happen again?: essays on instability and finance}.
\newblock Routledge, 2015.

\bibitem{haldane2011systemic}
Andrew~G Haldane and Robert~M May.
\newblock Systemic risk in banking ecosystems.
\newblock {\em Nature}, 469(7330):351--355, 2011.

\bibitem{gai2010contagion}
Prasanna Gai and Sujit Kapadia.
\newblock Contagion in financial networks.
\newblock {\em Proceedings of the Royal Society A: Mathematical, Physical and Engineering Sciences}, 466(2120):2401--2423, 2010.

\bibitem{squartini2013early}
Tiziano Squartini, Iman Van~Lelyveld, and Diego Garlaschelli.
\newblock Early-warning signals of topological collapse in interbank networks.
\newblock {\em Scientific reports}, 3(1):1--9, 2013.

\bibitem{caccioli2018network}
Fabio Caccioli, Paolo Barucca, and Teruyoshi Kobayashi.
\newblock Network models of financial systemic risk: a review.
\newblock {\em Journal of Computational Social Science}, 1:81--114, 2018.

\bibitem{carvalho}
Vasco~M Carvalho and Alireza Tahbaz-Salehi.
\newblock Production networks: A primer.
\newblock {\em Annual Review of Economics}, 11:635--663, 2019.

\bibitem{long1983real}
John~B Long~Jr and Charles~I Plosser.
\newblock Real business cycles.
\newblock {\em Journal of political Economy}, 91(1):39--69, 1983.

\bibitem{atalay2011network}
Enghin Atalay, Ali Hortacsu, James Roberts, and Chad Syverson.
\newblock Network structure of production.
\newblock {\em Proceedings of the National Academy of Sciences}, 108(13):5199--5202, 2011.

\bibitem{moran2019may}
Jos{\'e} Moran and Jean-Philippe Bouchaud.
\newblock May's instability in large economies.
\newblock {\em Physical Review E}, 100(3):032307, 2019.

\bibitem{hawkins1949note}
David Hawkins and Herbert~A Simon.
\newblock Note: some conditions of macroeconomic stability.
\newblock {\em Econometrica, Journal of the Econometric Society}, pages 245--248, 1949.

\bibitem{flaschel2010classical}
Peter Flaschel.
\newblock {\em The classical growth cycle: reformulation, simulation and some facts}.
\newblock Springer, 2010.

\bibitem{herskovic2020firm}
Bernard Herskovic, Bryan Kelly, Hanno Lustig, and Stijn Van~Nieuwerburgh.
\newblock Firm volatility in granular networks.
\newblock {\em Journal of Political Economy}, 128(11):4097--4162, 2020.

\bibitem{mazzarisi2024beyond}
Onofrio Mazzarisi and Matteo Smerlak.
\newblock Beyond may: Complexity-stability relationships in disordered dynamical systems.
\newblock {\em arXiv preprint arXiv:2403.11014}, 2024.

\bibitem{castellano2017relating}
Claudio Castellano and Romualdo Pastor-Satorras.
\newblock Relating topological determinants of complex networks to their spectral properties: Structural and dynamical effects.
\newblock {\em Physical Review X}, 7(4):041024, 2017.

\bibitem{perotti2009emergent}
Juan~I Perotti, Orlando~V Billoni, Francisco~A Tamarit, Dante~R Chialvo, and Sergio~A Cannas.
\newblock Emergent self-organized complex network topology out of stability constraints.
\newblock {\em Physical review letters}, 103(10):108701, 2009.

\bibitem{fisher1989disequilibrium}
Franklin~M Fisher.
\newblock Disequilibrium foundations of equilibrium economics.
\newblock {\em Cambridge Books}, 1989.

\bibitem{kaplan2018monetary}
Greg Kaplan, Benjamin Moll, and Giovanni~L Violante.
\newblock Monetary policy according to hank.
\newblock {\em American Economic Review}, 108(3):697--743, 2018.

\bibitem{baqaee2019macroeconomic}
David~Rezza Baqaee and Emmanuel Farhi.
\newblock The macroeconomic impact of microeconomic shocks: Beyond hulten's theorem.
\newblock {\em Econometrica}, 87(4):1155--1203, 2019.

\bibitem{baqaee2020productivity}
David~Rezza Baqaee and Emmanuel Farhi.
\newblock Productivity and misallocation in general equilibrium.
\newblock {\em The Quarterly Journal of Economics}, 135(1):105--163, 2020.

\bibitem{farmer2024making}
J~Doyne Farmer.
\newblock {\em Making Sense of Chaos: A Better Economics for a Better World}.
\newblock Yale University Press, 2024.

\bibitem{dessertaine2022out}
Th{\'e}o Dessertaine, Jos{\'e} Moran, Michael Benzaquen, and Jean-Philippe Bouchaud.
\newblock Out-of-equilibrium dynamics and excess volatility in firm networks.
\newblock {\em Journal of Economic Dynamics and Control}, 138:104362, 2022.

\bibitem{colon2022radical}
Celian Colon and Jean-Philippe Bouchaud.
\newblock The radical complexity of rewiring supplier--buyer networks.
\newblock {\em Available at SSRN 4300311}, 2022.

\bibitem{Sterman1989}
J.~Sterman.
\newblock Modeling managerial behavior: Misperceptions of feedback in a dynamic decision making experiment.
\newblock {\em Management Science}, 35(3):321--339, 1989.

\bibitem{nirei2024repricing}
Makoto Nirei and Jos{\'e}~A Scheinkman.
\newblock Repricing avalanches.
\newblock {\em Journal of Political Economy}, 132(4):1327--1388, 2024.

\bibitem{engle1982autoregressive}
Robert~F Engle.
\newblock Autoregressive conditional heteroscedasticity with estimates of the variance of united kingdom inflation.
\newblock {\em Econometrica: Journal of the econometric society}, pages 987--1007, 1982.

\bibitem{engle2001garch}
Robert Engle.
\newblock Garch 101: The use of arch/garch models in applied econometrics.
\newblock {\em Journal of economic perspectives}, 15(4):157--168, 2001.

\bibitem{leal2021repricing}
Laura Leal, Haaris Mateen, Makoto Nirei, and Jos{\'e}~A Scheinkman.
\newblock Repricing avalanches in the billion-prices data.
\newblock Technical report, National Bureau of Economic Research, 2021.

\bibitem{gualdi2015endogenous}
Stanislao Gualdi, Jean-Philippe Bouchaud, Giulia Cencetti, Marco Tarzia, and Francesco Zamponi.
\newblock Endogenous crisis waves: stochastic model with synchronized collective behavior.
\newblock {\em Physical review letters}, 114(8):088701, 2015.

\bibitem{gualdi2015tipping}
Stanislao Gualdi, Marco Tarzia, Francesco Zamponi, and Jean-Philippe Bouchaud.
\newblock Tipping points in macroeconomic agent-based models.
\newblock {\em Journal of Economic Dynamics and Control}, 50:29--61, 2015.

\bibitem{lombardi2021long}
Fabrizio Lombardi, Oren Shriki, Hans~J Herrmann, and Lucilla de~Arcangelis.
\newblock Long-range temporal correlations in the broadband resting state activity of the human brain revealed by neuronal avalanches.
\newblock {\em Neurocomputing}, 461:657--666, 2021.

\bibitem{sharma2021good}
Dhruv Sharma, Jean-Philippe Bouchaud, Marco Tarzia, and Francesco Zamponi.
\newblock Good speciation and endogenous business cycles in a constraint satisfaction macroeconomic model.
\newblock {\em Journal of Statistical Mechanics: Theory and Experiment}, 2021(6):063403, 2021.

\bibitem{bouchaud2001power}
Jean-Philippe Bouchaud.
\newblock Power laws in economics and finance: some ideas fromphysics.
\newblock {\em Quantitative finance}, 1(1):105, 2001.

\bibitem{bouchaud2011endogenous}
Jean-Philippe Bouchaud.
\newblock The endogenous dynamics of markets: Price impact, feedback loops and instabilities.
\newblock {\em Lessons from the credit crisis}, pages 345--74, 2011.

\bibitem{bouchaud2017have}
Jean-Philippe Bouchaud and Damien Challet.
\newblock Why have asset price properties changed so little in 200 years.
\newblock In {\em Econophysics and Sociophysics: Recent Progress and Future Directions}, pages 3--17. Springer, 2017.

\bibitem{muzy2013random}
Jean-Fran{\c{c}}ois Muzy, Rachel Ba{\"\i}le, and Emmanuel Bacry.
\newblock Random cascade model in the limit of infinite integral scale as the exponential of a nonstationary 1/f noise: Application to volatility fluctuations in stock markets.
\newblock {\em Physical Review E—Statistical, Nonlinear, and Soft Matter Physics}, 87(4):042813, 2013.

\bibitem{bayer2023rough}
Christian Bayer, Peter~K Friz, Masaaki Fukasawa, Jim Gatheral, Antoine Jacquier, and Mathieu Rosenbaum.
\newblock {\em Rough volatility}.
\newblock SIAM, 2023.

\bibitem{hardiman2014branching}
Stephen~J Hardiman and Jean-Philippe Bouchaud.
\newblock Branching-ratio approximation for the self-exciting hawkes process.
\newblock {\em Physical Review E}, 90(6):062807, 2014.

\bibitem{bacry2015hawkes}
Emmanuel Bacry, Iacopo Mastromatteo, and Jean-Fran{\c{c}}ois Muzy.
\newblock Hawkes processes in finance.
\newblock {\em Market Microstructure and Liquidity}, 1(01):1550005, 2015.

\bibitem{fair2002events}
Ray~C Fair.
\newblock Events that shook the market.
\newblock {\em The Journal of Business}, 75(4):713--731, 2002.

\bibitem{aubrun2024riding}
Cecilia Aubrun, Rudy Morel, Michael Benzaquen, and Jean-Philippe Bouchaud.
\newblock Riding wavelets: A method to discover new classes of price jumps.
\newblock {\em arXiv preprint arXiv:2404.16467}, 2024.

\bibitem{kyle1985continuous}
Albert~S Kyle.
\newblock Continuous auctions and insider trading.
\newblock {\em Econometrica: Journal of the Econometric Society}, pages 1315--1335, 1985.

\bibitem{bouchaud2000wealth}
Jean-Philippe Bouchaud and Marc M{\'e}zard.
\newblock Wealth condensation in a simple model of economy.
\newblock {\em Physica A: Statistical Mechanics and its Applications}, 282(3-4):536--545, 2000.

\bibitem{lillo2005theory}
Fabrizio Lillo, Szabolcs Mike, and J~Doyne Farmer.
\newblock Theory for long memory in supply and demand.
\newblock {\em Physical Review E—Statistical, Nonlinear, and Soft Matter Physics}, 71(6):066122, 2005.

\bibitem{bouchaud2009markets}
Jean-Philippe Bouchaud, J~Doyne Farmer, and Fabrizio Lillo.
\newblock How markets slowly digest changes in supply and demand.
\newblock In {\em Handbook of financial markets: dynamics and evolution}, pages 57--160. Elsevier, 2009.

\bibitem{bouchaud2018trades}
Jean-Philippe Bouchaud, Julius Bonart, Jonathan Donier, and Martin Gould.
\newblock {\em Trades, quotes and prices: financial markets under the microscope}.
\newblock Cambridge University Press, 2018.

\bibitem{sato2023inferring}
Yuki Sato and Kiyoshi Kanazawa.
\newblock Inferring microscopic financial information from the long memory in market-order flow: A quantitative test of the lillo-mike-farmer model.
\newblock {\em Physical Review Letters}, 131(19):197401, 2023.

\bibitem{farmer2004origin}
J~Doyne Farmer, Fabrizio Lillo, et~al.
\newblock On the origin of power-law tails in price fluctuations.
\newblock {\em Quantitative Finance}, 4(1):7--11, 2004.

\bibitem{gillemot2006there}
Laszlo Gillemot, J~Doyne Farmer, and Fabrizio Lillo.
\newblock There's more to volatility than volume.
\newblock {\em Quantitative finance}, 6(5):371--384, 2006.

\bibitem{fosset2020endogenous}
Antoine Fosset, Jean-Philippe Bouchaud, and Michael Benzaquen.
\newblock Endogenous liquidity crises.
\newblock {\em Journal of Statistical Mechanics: Theory and Experiment}, 2020(6):063401, 2020.

\bibitem{bochud2007optimal}
Thierry Bochud and Damien Challet.
\newblock Optimal approximations of power laws with exponentials: application to volatility models with long memory.
\newblock {\em Quantitative Finance}, 7(6):585--589, 2007.

\bibitem{zumbach2001heterogeneous}
Gilles Zumbach and Paul Lynch.
\newblock Heterogeneous volatility cascade in financial markets.
\newblock {\em Physica A: Statistical Mechanics and its Applications}, 298(3-4):521--529, 2001.

\bibitem{Wehrli2022}
Alexander Wehrli and Didier Sornette.
\newblock The excess volatility puzzle explained by financial noise amplification from endogenous feedbacks.
\newblock {\em Scientific Reports}, 12:18895, 2022.

\bibitem{bremaud2001hawkes}
Pierre Br{\'e}maud and Laurent Massouli{\'e}.
\newblock Hawkes branching point processes without ancestors.
\newblock {\em Journal of applied probability}, 38(1):122--135, 2001.

\bibitem{chicheportiche2014fine}
R{\'e}my Chicheportiche and Jean-Philippe Bouchaud.
\newblock The fine-structure of volatility feedback i: Multi-scale self-reflexivity.
\newblock {\em Physica A: Statistical Mechanics and its Applications}, 410:174--195, 2014.

\bibitem{hardiman2013critical}
Stephen~J Hardiman, Nicolas Bercot, and Jean-Philippe Bouchaud.
\newblock Critical reflexivity in financial markets: a hawkes process analysis.
\newblock {\em The European Physical Journal B}, 86:1--9, 2013.

\bibitem{filimonov2014quantification}
Vladimir Filimonov, David Bicchetti, Nicolas Maystre, and Didier Sornette.
\newblock Quantification of the high level of endogeneity and of structural regime shifts in commodity markets.
\newblock {\em Journal of international Money and finance}, 42:174--192, 2014.

\bibitem{blanc2017quadratic}
Pierre Blanc, Jonathan Donier, and J-P Bouchaud.
\newblock Quadratic hawkes processes for financial prices.
\newblock {\em Quantitative Finance}, 17(2):171--188, 2017.

\bibitem{wheatley2019endo}
Spencer Wheatley, Alexander Wehrli, and Didier Sornette.
\newblock The endo--exo problem in high frequency financial price fluctuations and rejecting criticality.
\newblock {\em Quantitative Finance}, 19(7):1165--1178, 2019.

\bibitem{wyart2008relation}
Matthieu Wyart, Jean-Philippe Bouchaud, Julien Kockelkoren, Marc Potters, and Michele Vettorazzo.
\newblock Relation between bid--ask spread, impact and volatility in order-driven markets.
\newblock {\em Quantitative finance}, 8(1):41--57, 2008.

\bibitem{glosten1985bid}
Lawrence~R Glosten and Paul~R Milgrom.
\newblock Bid, ask and transaction prices in a specialist market with heterogeneously informed traders.
\newblock {\em Journal of financial economics}, 14(1):71--100, 1985.

\bibitem{dall2019does}
Lorenzo Dall’Amico, Antoine Fosset, Jean-Philippe Bouchaud, and Michael Benzaquen.
\newblock How does latent liquidity get revealed in the limit order book?
\newblock {\em Journal of Statistical Mechanics: Theory and Experiment}, 2019(1):013404, 2019.

\bibitem{fosset2022non}
Antoine Fosset, Jean-Philippe Bouchaud, and Michael Benzaquen.
\newblock Non-parametric estimation of quadratic hawkes processes for order book events.
\newblock {\em The European Journal of Finance}, 28(7):663--678, 2022.

\bibitem{bouchaud2003fluctuations}
Jean-Philippe Bouchaud, Yuval Gefen, Marc Potters, and Matthieu Wyart.
\newblock Fluctuations and response in financial markets: thesubtle nature ofrandom'price changes.
\newblock {\em Quantitative finance}, 4(2):176, 2003.

\bibitem{farmer2002market}
J~Doyne Farmer.
\newblock Market force, ecology and evolution.
\newblock {\em Industrial and Corporate Change}, 11(5):895--953, 2002.

\bibitem{lux1999scaling}
Thomas Lux and Michele Marchesi.
\newblock Scaling and criticality in a stochastic multi-agent model of a financial market.
\newblock {\em Nature}, 397(6719):498--500, 1999.

\bibitem{scholl2021market}
Maarten~P Scholl, Anisoara Calinescu, and J~Doyne Farmer.
\newblock How market ecology explains market malfunction.
\newblock {\em Proceedings of the National Academy of Sciences}, 118(26):e2015574118, 2021.

\bibitem{chiarella1992dynamics}
Carl Chiarella.
\newblock The dynamics of speculative behaviour.
\newblock {\em Annals of operations research}, 37(1):101--123, 1992.

\bibitem{palmer1994artificial}
Richard~G Palmer, W~Brian Arthur, John~H Holland, Blake LeBaron, and Paul Tayler.
\newblock Artificial economic life: a simple model of a stockmarket.
\newblock {\em Physica D: Nonlinear Phenomena}, 75(1-3):264--274, 1994.

\bibitem{brock1998heterogeneous}
William~A Brock and Cars~H Hommes.
\newblock Heterogeneous beliefs and routes to chaos in a simple asset pricing model.
\newblock {\em Journal of Economic dynamics and Control}, 22(8-9):1235--1274, 1998.

\bibitem{giardina2003bubbles}
Irene Giardina and J-P Bouchaud.
\newblock Bubbles, crashes and intermittency in agent based market models.
\newblock {\em The European Physical Journal B-Condensed Matter and Complex Systems}, 31:421--437, 2003.

\bibitem{bouchaud2017black}
J.~P. Bouchaud, S.~Ciliberti, Y.~Lempériere, A.~Majewski, P.~Seager, and K.~S. Ronia.
\newblock Black was right: Price is within a factor 2 of value.
\newblock {\em arXiv preprint arXiv:1711.04717}, 2017.

\bibitem{majewski2020}
Adam~A. Majewski, Stefano Ciliberti, and Jean-Philippe Bouchaud.
\newblock Co-existence of trend and value in financial markets: Estimating an extended chiarella model.
\newblock {\em Journal of Economic Dynamics and Control}, 112:103791, 2020.

\bibitem{schmitt2021trend}
Noemi Schmitt and Frank Westerhoff.
\newblock Trend followers, contrarians and fundamentalists: Explaining the dynamics of financial markets.
\newblock {\em Journal of Economic Behavior \& Organization}, 192:117--136, 2021.

\bibitem{lux2021can}
Thomas Lux.
\newblock Can heterogeneous agent models explain the alleged mispricing of the s\&p 500?
\newblock {\em Quantitative Finance}, 21(9):1413--1433, 2021.

\bibitem{challet2004minority}
Damien Challet, Matteo Marsili, and Yi-Cheng Zhang.
\newblock {\em Minority games: interacting agents in financial markets}.
\newblock OUP Oxford, 2004.

\bibitem{challet2003criticality}
Damien Challet and Matteo Marsili.
\newblock Criticality and market efficiency in a simple realistic model of the stock market.
\newblock {\em Physical Review E}, 68(3):036132, 2003.

\bibitem{alfi2009self}
Valentina Alfi, Luciano Pietronero, and A~Zaccaria.
\newblock Self-organization for the stylized facts and finite-size effects in a financial-market model.
\newblock {\em Europhysics Letters}, 86(5):58003, 2009.

\bibitem{grossman1980impossibility}
Sanford~J Grossman and Joseph~E Stiglitz.
\newblock On the impossibility of informationally efficient markets.
\newblock {\em The American economic review}, 70(3):393--408, 1980.

\bibitem{amir2005market}
Rabah Amir, Igor Evstigneev, Thorsten Hens, and Klaus Schenk-Hoppé.
\newblock Market selection and survival of investment strategies.
\newblock {\em Journal of Mathematical Economics}, 41(1-2):105--122, 2005.

\bibitem{caccioli2009eroding}
Fabio Caccioli, Matteo Marsili, and Pierpaolo Vivo.
\newblock Eroding market stability by proliferation of financial instruments.
\newblock {\em The European Physical Journal B}, 71:467--479, 2009.

\bibitem{marsili2014complexity}
Matteo Marsili.
\newblock Complexity and financial stability in a large random economy.
\newblock {\em Quantitative Finance}, 14(9):1663--1675, 2014.

\bibitem{cherkashin2009reality}
Dmitriy Cherkashin, J~Doyne Farmer, and Seth Lloyd.
\newblock The reality game.
\newblock {\em Journal of Economic Dynamics and Control}, 33(5):1091--1105, 2009.

\bibitem{caccioli2012impact}
Fabio Caccioli, Jean-Philippe Bouchaud, and D~Farmer.
\newblock Impact-adjusted valuation and the criticality of leverage.
\newblock {\em Risk}, 25(12):74--77, 2012.

\bibitem{cont2013running}
Rama Cont and Lakshithe Wagalath.
\newblock Running for the exit: distressed selling and endogenous correlation in financial markets.
\newblock {\em Mathematical Finance: An International Journal of Mathematics, Statistics and Financial Economics}, 23(4):718--741, 2013.

\bibitem{caccioli2015overlapping}
Fabio Caccioli, J~Doyne Farmer, Nick Foti, and Daniel Rockmore.
\newblock Overlapping portfolios, contagion, and financial stability.
\newblock {\em Journal of Economic Dynamics and Control}, 51:50--63, 2015.

\bibitem{khandani2007happened}
Amir~E Khandani and Andrew~W Lo.
\newblock What happened to the quants in august 2007? {E}vidence from factors and transactions data.
\newblock {\em Journal of Financial Markets}, 14(1):1--46, 2011.

\bibitem{moran2023temporal}
Jos{\'e} Moran, Frank~P Pijpers, Utz Weitzel, Jean-Philippe Bouchaud, and Debabrata Panja.
\newblock Temporal criticality in socio-technical systems.
\newblock {\em arXiv preprint arXiv:2307.03546}, 2023.

\bibitem{taleb}
Nassim~Nicholas Taleb.
\newblock {\em Antifragile: Things that gain from disorder}, volume~3.
\newblock Random House Trade Paperbacks, 2014.

\bibitem{Hynes2022}
W.~Hynes, D~Trump, B. D.~Kirman, A.~Haldane, and I.~Linkov.
\newblock Systemic resilience in economics.
\newblock {\em Nature Physics}, 18:381--384, 2022.

\end{thebibliography}

\end{document}